%% file: ms.tex
\documentclass[preprint]{aastex}
\received{}
\accepted{}
\newcommand{\monolum}{\mbox{\,erg\,s$^{-1}$\,Hz$^{-1}$}}

\slugcomment{24 January 2006 Draft}

\author{
Gordon T. Richards,\altaffilmark{1,2}
Mark Lacy,\altaffilmark{3}
Lisa J. Storrie-Lombardi,\altaffilmark{3}
Patrick B. Hall,\altaffilmark{4}
S. C. Gallagher,\altaffilmark{5}
Dean C. Hines,\altaffilmark{6}
Xiaohui Fan,\altaffilmark{7}
Casey Papovich,\altaffilmark{7}
Daniel E. Vanden Berk,\altaffilmark{8}
George B. Trammell,\altaffilmark{8}
Donald P. Schneider,\altaffilmark{8}
Marianne Vestergaard,\altaffilmark{7}
Donald G. York,\altaffilmark{9,10}
Sebastian Jester,\altaffilmark{11,12}
Scott F. Anderson,\altaffilmark{13}
Tam\'as Budav\'ari,\altaffilmark{2}
and Alexander S. Szalay\altaffilmark{2}
}

\altaffiltext{1}{Princeton University Observatory, Peyton Hall, Princeton, NJ 08544.}
\altaffiltext{2}{Department of Physics and Astronomy, The Johns Hopkins University, 3400 North Charles Street, Baltimore, MD 21218-2686.}
\altaffiltext{3}{Spitzer Science Center, Caltech, Mail Code 220-6, Pasadena, CA 91125.}
\altaffiltext{4}{Department of Physics and Astronomy, York University, 4700 Keele Street, Toronto, Ontario, M3J 1P3, Canada.}
\altaffiltext{5}{Department of Physics and Astronomy, UCLA, Mail Code 154705, 475 Portola Plaza, Los Angeles, CA 90095.}
\altaffiltext{6}{Space Science Institute, 4750 Walnut Street, Suite 205, Boulder, CO 80301.}
\altaffiltext{7}{Steward Observatory, University of Arizona, 933 North Cherry Avenue, Tucson, AZ 85721.}
\altaffiltext{8}{Department of Astronomy and Astrophysics, The Pennsylvania State University, 525 Davey Laboratory, University Park, PA 16802.}
\altaffiltext{9}{Department of Astronomy and Astrophysics, The University of Chicago, 5640 South Ellis Avenue, Chicago, IL 60637.}
\altaffiltext{10}{Enrico Fermi Institute, The University of Chicago, 5640 South Ellis Avenue, Chicago, IL 60637.}
\altaffiltext{11}{Fermi National Accelerator Laboratory, P.O. Box 500, Batavia, IL 60510.}
\altaffiltext{12}{School of Physics and Astronomy, Southampton University, Southampton SO17 1BJ, UK.}
\altaffiltext{13}{Department of Astronomy, University of Washington, Box 351580, Seattle, WA 98195.}

\begin{document}

%\title{The Mid-Infrared/Optical Properties of Type 1 Quasars}
\title{Spectral Energy Distributions and Multiwavelength Selection of Type 1 Quasars}

\begin{abstract}

We present an analysis of the mid-infrared (MIR) and optical
properties of type 1 (broad-line) quasars detected by the {\em Spitzer
Space Telescope}.  The MIR color-redshift relation is characterized to
$z\sim3$, with predictions to $z=7$.  We demonstrate how combining MIR
and optical colors can yield even more efficient selection of active
galactic nuclei (AGN) than MIR or optical colors alone.  Composite
spectral energy distributions (SEDs) are constructed for 259 quasars
with both Sloan Digital Sky Survey and {\em Spitzer} photometry,
supplemented by near-IR, {\em GALEX}, VLA and {\em ROSAT} data where
available.  We discuss how the spectral diversity of quasars
influences the determination of bolometric luminosities and accretion
rates; assuming the mean SED can lead to errors as large as a factor
of 2 for individual quasars.  Finally, we show that careful
consideration of the shape of the mean quasar SED and its redshift
dependence leads to a lower estimate of the fraction of
reddened/obscured AGNs missed by optical surveys as compared to
estimates derived from a single mean MIR to optical flux ratio.

\end{abstract}


\section{Introduction}

Access to the mid-infrared (MIR) region opens up new realms for quasar
science as we are able to study large numbers of objects with high
signal-to-noise ratio data in this bolometrically important band for
the first time.  At least four distinct energy generation mechanisms
are at work in active galactic nuclei (AGN) from jets in the radio,
dust in the IR, accretion disks in the optical-UV-soft-Xray,
and Compton upscattering in the hard-X-ray.  All of these spectral
regions need to be sampled with high precision if we are to understand
the physical processes governing AGN emission.  The {\em Spitzer Space
Telescope} \markcite{wrl+04}({Werner} {et~al.} 2004) allows the first robust glimpse of the
physics of the putative dusty torus in AGNs out to $z\sim2$--3.

MIR photometry from {\em Spitzer} has provided a better census of
active nuclei in galaxies than has been previously possible
\markcite{lss+04}(e.g., {Lacy} {et~al.} 2004a).  Optical surveys are biased against heavily
reddened and obscured objects, and even X-ray surveys may fail to
uncover Compton-thick sources.  Thus the MIR presents an attractive
window for determining the black hole accretion history of the
Universe.  To that end, {\em Spitzer} will be of considerable utility
in helping to decipher the nature of the $M_{\rm BH}-\sigma$ relation
\markcite{tgb+02}(e.g., {Tremaine} {et~al.} 2002), both in terms of the census of AGNs and in
determining accurate bolometric luminosities (and, in turn, accretion
rates).

This paper builds upon and extends the results from recent papers
describing the {\em Spitzer} MIR color distribution of AGNs.
\markcite{lss+04}{Lacy} {et~al.} (2004a) showed that MIR colors alone can be used to select AGNs
with both high efficiency and completeness, including both dust
reddened and optically obscured (type 2) AGNs that may otherwise be
overlooked by optical selection techniques.  We will show that the
addition of optical colors and morphology can be used to improve the
MIR-only selection efficiency of type 1 quasars (including those that
are moderately reddened).

\markcite{seg+04}{Stern} {et~al.} (2005) also describe a MIR selection technique for AGN, making
statistical arguments that the obscured AGN fraction may be as high as
76\%.  We reconsider their arguement in light of the influence that
the shape of the mean quasar spectral energy distribution (SED) has on
determining the obscured quasar fraction.  Such considerations allow
us to demonstrate that the true obscured AGN fraction must be
lower than that determined by \markcite{seg+04}{Stern} {et~al.} (2005).  

Finally, \markcite{hpp+04}{Hatziminaoglou} {et~al.} (2005) investigated the combined optical+MIR color
distribution of quasars by combining data from the ELAIS-N1 field in
the {\em Spitzer} Wide-Area Infrared Extragalactic Survey (SWIRE;
\markcite{lsr+03}{Lonsdale} {et~al.} 2003) with data from the Sloan Digital Sky Survey (SDSS;
\markcite{yaa+00}{York} {et~al.} 2000).  Using the data from 35 SDSS quasars
%\markcite{rfn+02,shr+05}({Richards} {et~al.} 2002; {Schneider} {et~al.} 2005) 
they determine the mean optical-MIR SED of type 1 quasars and
investigate their mass and bolometric luminosity distribution.  We
expand on these results by determining a number of different ``mean''
SEDs as a function of color and luminosity for 259 SDSS quasars in the
{\em Spitzer} Extragalactic First Look
Survey\footnote{http://ssc.spitzer.caltech.edu/fls/} (XFLS),
SWIRE\footnote{http://swire.ipac.caltech.edu/swire/} ELAIS-N1/N2, and
SWIRE Lockman Hole areas.  We use these SEDs to demonstrate that the
diversity of quasar SEDs must be considered when determining
bolometric luminosities and accretion rates for individual quasars ---
as was emphasized in the seminal SED work of \markcite{ewm+94}{Elvis} {et~al.} (1994).

Section~2 reviews the data sets used in our analysis.  In \S~3 we
explore the MIR color-redshift relation and MIR-optical color-color
space occupied by type 1 quasars.  In addition to showing these
relations for the data, we also show the predicted relations derived
from two quasar SEDs convolved with the SDSS and {\em Spitzer} filters
curves: one SED derived largely from broad-band photometry
\markcite{ewm+94}({Elvis} {et~al.} 1994), the other from a mean optical+IR spectral template
\markcite{ghw05}({Glikman et al.} 2005).  Section~4 presents a brief discussion of the
determination of the type 1 to type 2 ratio of quasars. In \S~5 we
discuss the radio through X-ray SED of quasars and construct new
MIR-optical templates from our sample. We present an overall mean SED
along with mean SEDs for subsets of optically luminous/dim, MIR
luminous/dim, and optically blue/red quasars in order to explore how
different optical/MIR properties are related to the overall
SED. Section~6 discusses the implications of our new SED templates on
the determination of bolometric luminosities and accretion rates.  Our
conclusions are presented in \S~7.  

Throughout this paper we will distinguish between normal type 1
quasars, dust reddened/extincted type 1 quasars, and type 2 quasars.
By type 1 quasars we mean those quasars having broad lines and
colors/spectral indices that are roughly consistent with a Gaussian
spectral index distribution of $\alpha_{\nu}=-0.5\pm0.3$
($f_{\nu}\propto\nu^{\alpha}$).  Reddened type 1 quasars are those
quasars that have broad lines but have spectral indices that are
redder than about $\alpha_{\nu}=-1$.  Optical surveys can find such
quasars up to $E(B-V)\sim0.5$, but are increasingly incomplete above
$E(B-V)\sim0.1$ \markcite{rhv+03}({Richards} {et~al.} 2003).  By type 2 quasars, we mean those that
lack rest-frame optical/UV broad emission lines and with nuclei that
are completely obscured in the optical such that the optical colors
are consistent with the host galaxy.  Throughout this paper we use a
$\Lambda$-CDM cosmology with $H_{\rm o}=70$\,km\,s$^{-1}$\,Mpc$^{-1}$,
$\Omega_{\Lambda}=0.7$, and $\Omega_{\rm m}=0.3$, consistent with the
WMAP cosmology \markcite{svp+03}({Spergel} {et~al.} 2003).

\section{The Data}

We investigate the mid-IR and optical properties of type 1 quasars
that are detected in both the SDSS and in all four bands of the {\em
Spitzer} Infrared Array Camera (IRAC; \markcite{fha+04}{Fazio} {et~al.} 2004).  The {\em
Spitzer} data are taken from the XFLS and
SWIRE ELAIS-N1,
ELAIS-N2, and Lockman Hole areas which have (RA [deg], Dec [deg])
centers of (259.5,59.5), (242.75,55.0), (249.2,41.029), and
(161.25,58.0), respectively.

We begin with SDSS type 1 quasars cataloged by \markcite{shr+05}{Schneider} {et~al.} (2005), the
majority of which were selected by the algorithm given by
\markcite{rfn+02}{Richards} {et~al.} (2002).  This catalog includes matches to the FIRST
\markcite{bwh95}({Becker}, {White}, \& {Helfand} 1995) survey with the VLA, ROSAT \markcite{vab+00}({Voges} {et~al.} 2000), and 2MASS
\markcite{sss+97}({Skrutskie} {et~al.} 1997).  For a definition of the SDSS photometric system, see
\markcite{fig+96}{Fukugita} {et~al.} (1996); \markcite{aaa+05}{Adelman-McCarthy} {et~al.} (2006) provide a description of the latest
SDSS data release (DR4).  All SDSS magnitudes have been corrected for
Galactic extinction according to \markcite{sfd98}{Schlegel}, {Finkbeiner}, \&  {Davis} (1998).

\begin{figure}
\epsscale{1.0}
%\plottwo{figures/radecfls.ps}{figures/radeclock.ps}
\plottwo{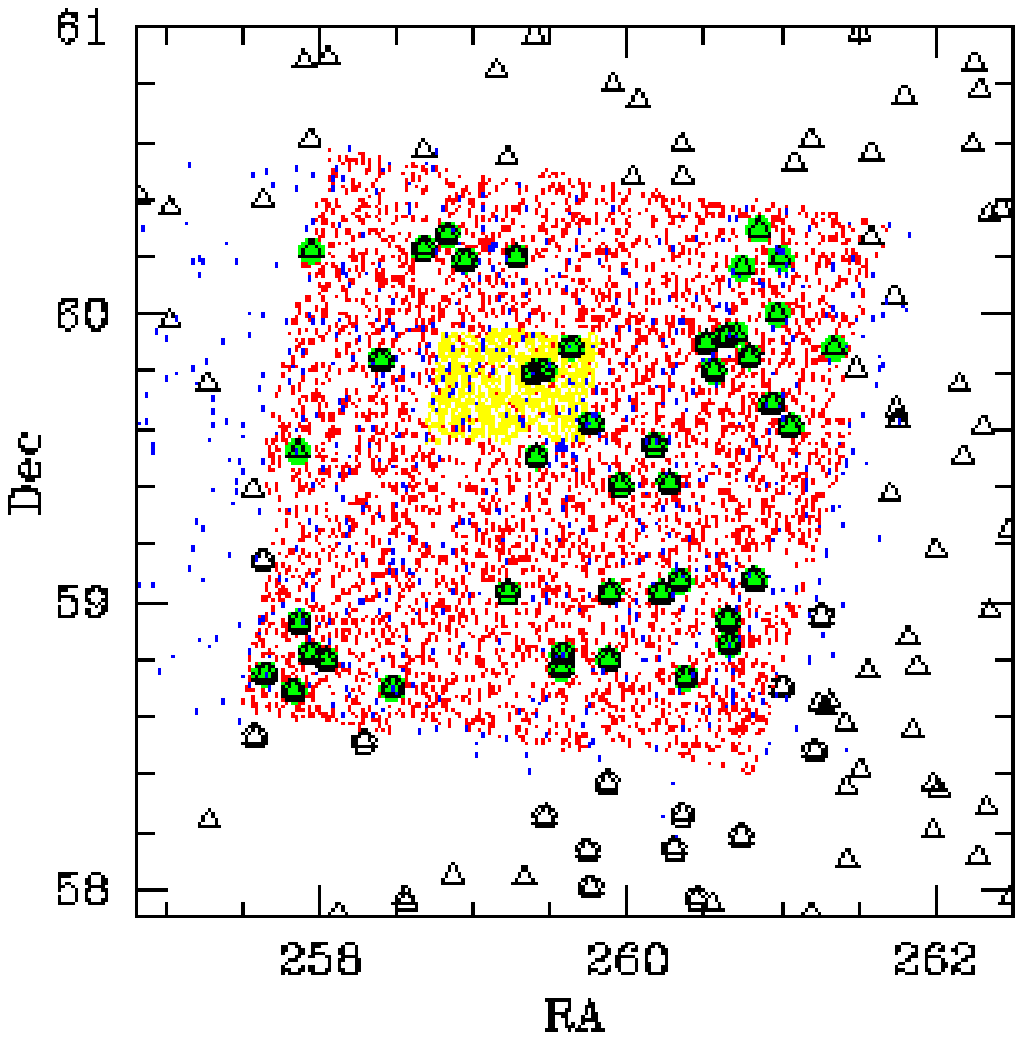}{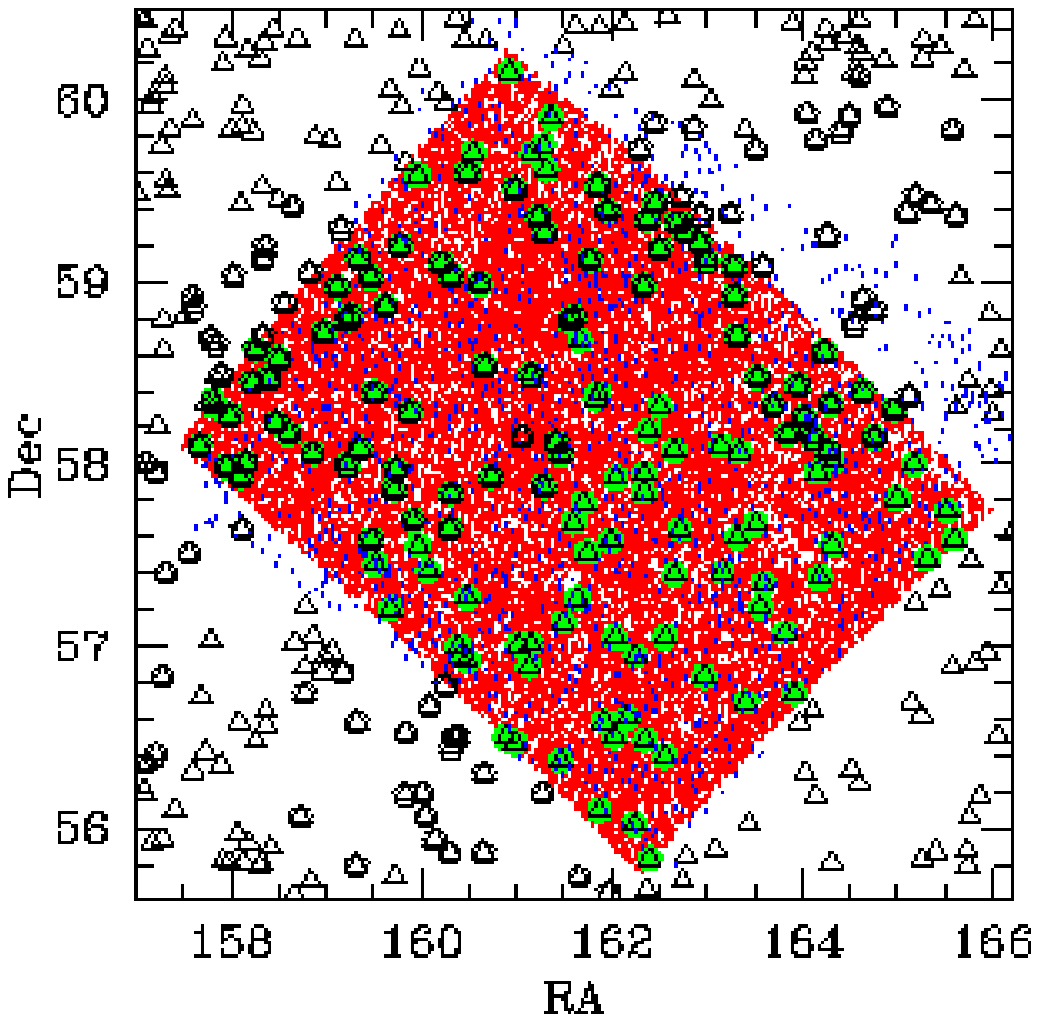}
\caption{Location of SDSS-DR3 quasars in the XFLS ({\em left}) and SWIRE
Lockman Hole ({\em right}) fields.  Red/yellow/blue points are
IRAC/IRACverification/MIPS70 sources.  Open triangles are
SDSS-DR3 quasars.  Green circles are SDSS-DR3 quasars with IRAC
detections in all 4 bands.  Open pentagons indicate {\em
GALEX}-detected SDSS quasars.
\label{fig:fig1}}
\end{figure}

\begin{figure}
\epsscale{1.0}
%\plottwo{figures/radecen1.ps}{figures/radecen2.ps}
\plottwo{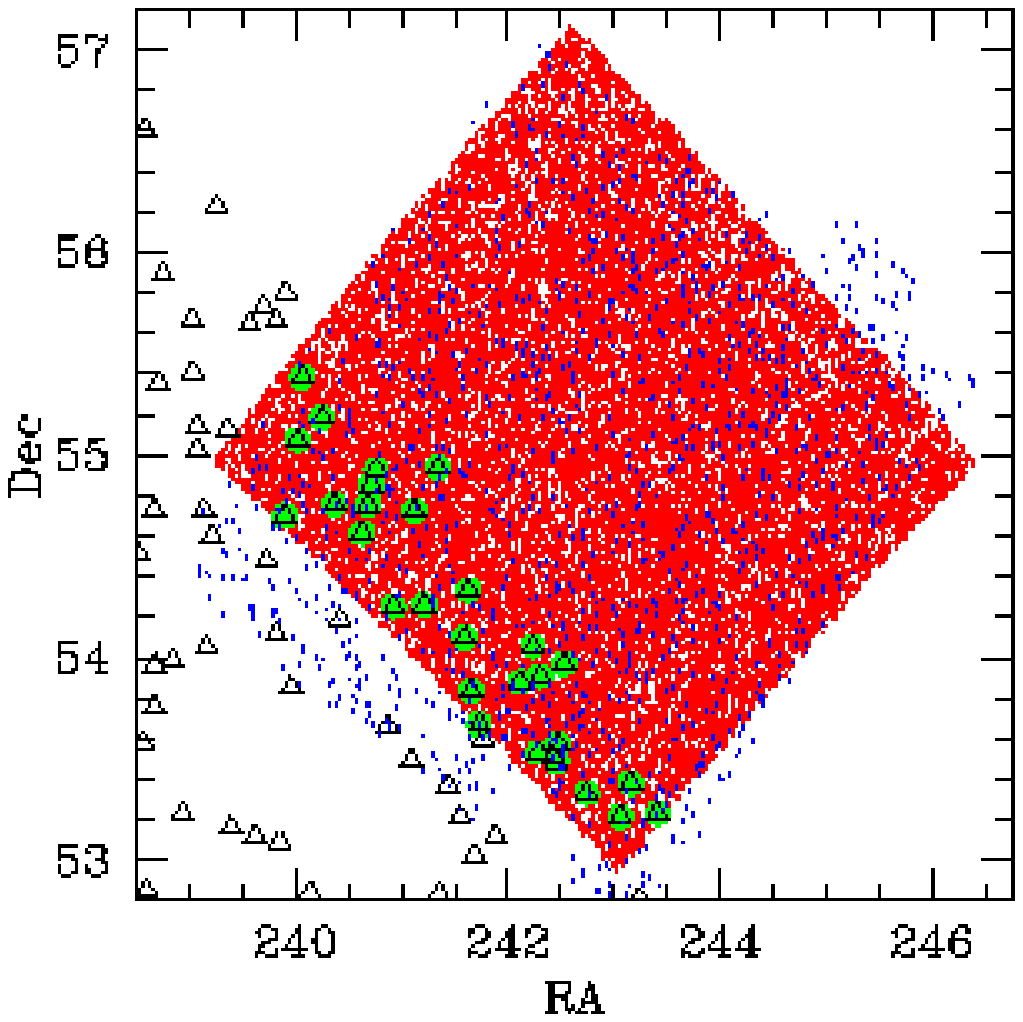}{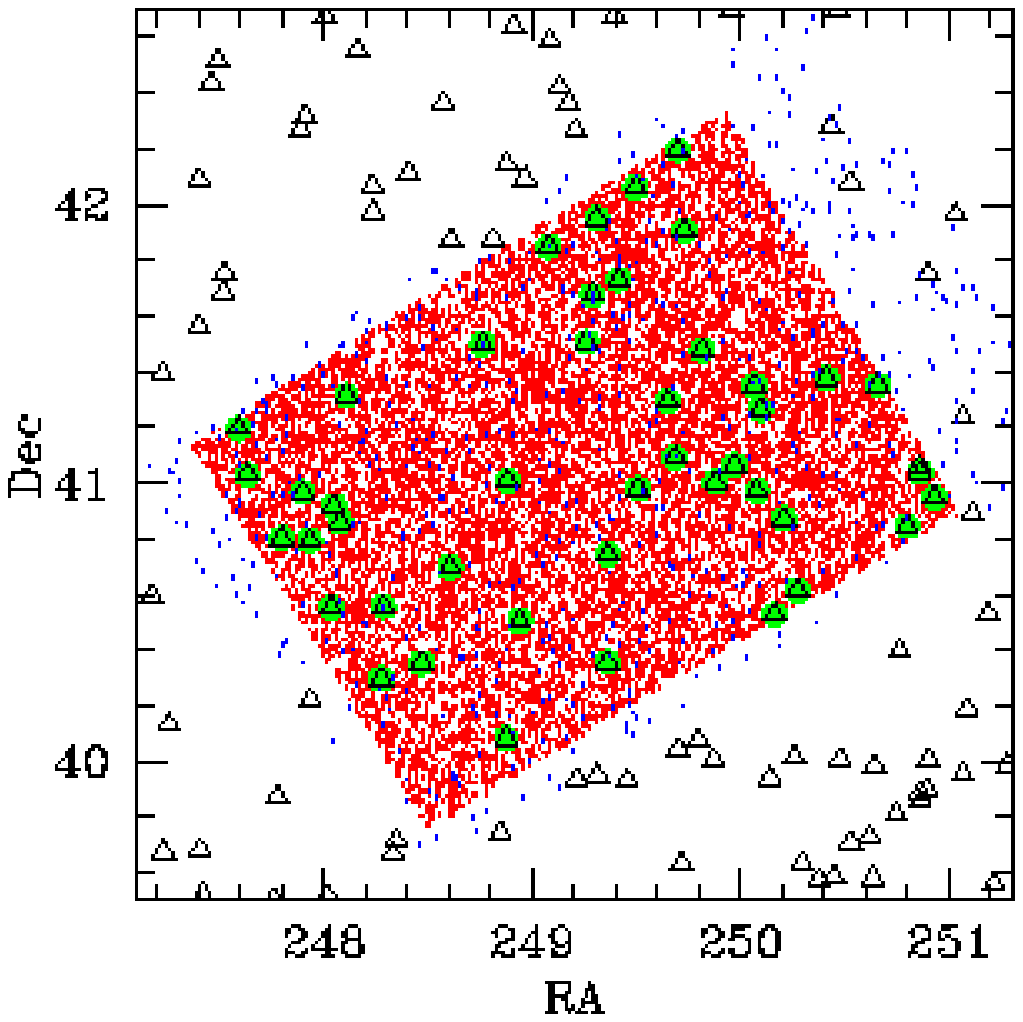}
\caption{Location of SDSS-DR3 quasars in the SWIRE ELAIS N1 ({\em
left}) and N2 ({\em right}) fields.  Red points are 4-band IRAC
sources.  Blue points are MIPS 70 $\mu$m sources.  Open triangles are
SDSS-DR3 quasars.  Green circles are SDSS-DR3 quasars with IRAC
detections in all 4 bands.
\label{fig:fig2}}
\end{figure}

\begin{sloppypar}
The 46,420 SDSS quasars of \markcite{shr+05}{Schneider} {et~al.} (2005) are matched to IRAC
detections in the XFLS \markcite{lacy05}(main\_4band.cat; {Lacy} {et~al.} 2005) and the
SWIRE ELAIS-N1, -N2, and Lockman Hole
(SWIRE2\_N1\_cat\_IRAC24\_16jun05.tbl,
SWIRE2\_N2\_cat\_IRAC24\_16jun05.tbl,
SWIRE2\_Lockman\_cat\_IRAC24\_10Nov05.tbl; Surace et al.\ 2005) areas
of sky.  The IRAC bandpasses are generally refereed to as Channels 1
through 4 or as the 3.6$\mu$m, 4.5$\mu$m, 5.8$\mu$m, and 8.0$\mu$m
bands, respectively.  For a quasar spectrum with $\alpha_{\nu}=-1$
($f_{\nu} \propto \nu^{\alpha}$), the effective wavelengths of the
IRAC bandpasses are actually closer to 3.52$\mu$m, 4.46$\mu$m,
5.67$\mu$m, and 7.70$\mu$m.  The SWIRE catalogs also include $24\mu$m
photometry from the Multiband Imaging Photometer for {\em Spitzer}
(MIPS; \markcite{rye+04}{Rieke} {et~al.} 2004).  In the XFLS field, 24$\mu$m sources are
cataloged by Fadda et al. (2006) and we include matches from that
catalog as well.  As the limits of the mid-IR catalogs are much deeper
than the SDSS spectroscopic survey, we consider only objects detected
in all 4 IRAC bands.  Within a matching radius of $1\farcs0$ there are
44 SDSS-DR3 quasar matches in the XFLS area, 29 in the ELAIS-N1 area,
44 in the ELAIS-N2 area, and 142 in the Lockman Hole area.  All but
one of the optically selected SDSS quasars has 4-band IRAC coverage in
the regions of overlap between the SDSS and {\em Spitzer} data; see
Figures~\ref{fig:fig1} and ~\ref{fig:fig2}.  The exception being
SDSSJ~104413.47+580858.9 ($z=3.7$) which has only a limit in IRAC
channel 3.
\end{sloppypar}

To construct the most detailed quasar spectral energy distributions
(SEDs) possible, we include data available at other wavelengths.
%We include $24\mu$m detections for XFLS sources from the catalog of
%\markcite{fadda06} ().  
We include matches to MIPS 70$\mu$m sources in the XFLS
(FLS70\_sn7\_jul05.txt; \markcite{ffy+05}{Frayer} {et~al.} 2006) and in the SWIRE
(SWIRE2\_EN1\_70um\_23nov05.tbl, SWIRE2\_EN2\_70um\_23nov05.tbl,
SWIRE3\_Lockman\_70um\_23nov05.tbl; Surace et al.\ 2005) areas.  No
MIPS 160$\mu$m data are included as the flux density limits of these
data in the XFLS and SWIRE areas are much brighter than expected flux
densities of even the brightest SDSS-DR3 quasars in these fields.  For
the SDSS quasars in the ELAIS fields we have extracted $15\mu m$
photometry from the \markcite{rlp+04}{Rowan-Robinson} {et~al.} (2004) catalog.  We also extract $J/H/K$
and radio information from this catalog if that information was not
otherwise available.

Some of these areas of sky have been observed by {\em GALEX}
\markcite{mfs+05}({Martin} {et~al.} 2005), and the data were released as part of {\em GALEX}-GR1.
Quasars are readily detected by {\em GALEX} (see \markcite{bsz+05}{Bianchi} {et~al.} 2005 and
\markcite{sbr+05}{Seibert} {et~al.} 2005), thus we also include {\em GALEX} photometry where
available.  Matching of the {\em GALEX} catalogs and the SDSS DR3
quasar sample is described by \markcite{tvs05}{Trammell} {et~al.} (2005).  The effective
wavelengths of the {\em GALEX} NUV and FUV bandpasses (hereafter
referred to as $n$ and $f$ magnitudes) are 1516\AA\ and 2267\AA.  {\em
GALEX} photometry has been corrected for Galactic extinction assuming
$A_{n}/E(B-V)=8.741$ and $A_{f}/E(B-V)=8.376$ \markcite{wtm+05}({Wyder} {et~al.} 2005).  A total
of 55 and 88 of the DR3 quasars have {\em GALEX} detections in the
$f$- and $n$-bands, respectively.

In the radio, we have matched to the deeper VLA data taken in the XFLS
area by \markcite{ccy+03}{Condon} {et~al.} (2003), which catalogs 5$\sigma$ detections with
fluxes higher than 115$\mu$Jy (about an order of magnitude deeper than
FIRST).  Deep VLA data also exists for the ELAIS and Lockman Hole
areas, but only over a small area of sky \markcite{cmm+99,czh+03}(e.g., {Ciliegi} {et~al.} 1999, 2003).

Most of our objects are fainter than the 2MASS \markcite{sss+97}({Skrutskie} {et~al.} 1997) limits,
but we have supplemental near-IR data for a few.  Near-IR ($JHK_s$)
magnitudes for SDSS J1716+5902 were obtained on UT 9 September 2003
using the GRIM II instrument on the Apache Point Observatory 3.5-m
telescope.  Dithered images were obtained and reduced in the standard
fashion, using running flat-fielding and sky-subtraction
\markcite{hgc98}(e.g., {Hall}, {Green}, \& {Cohen} 1998) with all available good images in a given
filter for each object.  Four other sources (SDSSJ~171732.94+594747.5,
SDSSJ~171736.90+593011.4, SDSSJ~171748.43+594820.6, and
SDSSJ~171831.73+595309.4) were observed at Palomar Observatory.

Finally, to better characterize the optical+MIR color distribution of
type 1 quasars, we include 87 broadline quasars that are fainter than
the SDSS spectroscopic magnitude limit, but that were confirmed from
Hectospec \markcite{ffr+05}({Fabricant} {et~al.} 2005) follow-up of MIPS 24 $\mu$m sources by
\markcite{pap+05}{Papovich et al.} (2005).

Tables~\ref{tab:tab1} and \ref{tab:tab2} present all of the
multiwavelength data for the SDSS-DR3 quasars in our sample.  The
columns in Table~\ref{tab:tab1} are name, redshift, bolometric
luminosity (see \S~\ref{sec:bol}), integrated optical luminosity,
integrated IR luminosity, bolometric correction (from rest-frame
5100\AA\ to the bolometric luminosity), X-ray ({\em ROSAT}) count
rate, {\em GALEX} $f$- and $n$-band magnitudes, SDSS $ugriz$
magnitudes.  The columns in Table~\ref{tab:tab2} are name (again),
$J/H/K$ magnitudes, {\em Spitzer}-IRAC 3.6, 4.5, 5.8, and 8.0$\mu$m
flux densities, {\em ISO} 15$\mu$m flux density, {\em Spitzer}-MIPS 24
and 70$\mu$m flux densities, radio (VLA, 20\,cm) flux density, and
radio luminosity.  Both Tables~\ref{tab:tab1} and \ref{tab:tab2}
report only the {\em observed} photometry of these sources; no host
galaxy contamination has been removed (\S~\ref{sec:meansed}).  Of
these 259 quasars, 28 have ROSAT detections and 30 have radio
detections.  Of those 30 radio detections, only 8 are radio loud
($L_{\rm rad}>10^{33}$\monolum).

%Of these 117 quasars, 17 have ROSAT detections and 18 have radio
%detections.  Of those 18 radio detections, 5 are radio loud ($L_{\rm
%rad}>10^{33}$\monolum).

\section{MIR/Optical Colors of Type 1 Quasars}

\markcite{lss+04}{Lacy} {et~al.} (2004a), \markcite{seg+04}{Stern} {et~al.} (2005), and \markcite{hpp+04}{Hatziminaoglou} {et~al.} (2005) have already
demonstrated that MIR selection of AGNs (both type 1 and type 2) from
{\em Spitzer} photometry is very efficient.  Here we demonstrate that
this capability can be enhanced by including morphology and optical
color information from SDSS.  We concentrate on relatively unobscured
type 1 quasars, but to the extent that dust-reddened type 1 quasars
and obscured type 2 quasars are not extincted beyond the SDSS flux
density limits, what we learn here will apply to those objects as well.

\subsection{The Color-Redshift Relations}

\begin{figure}
\epsscale{1.0}
%\plotone{figures/czplot2.ps}
\plotone{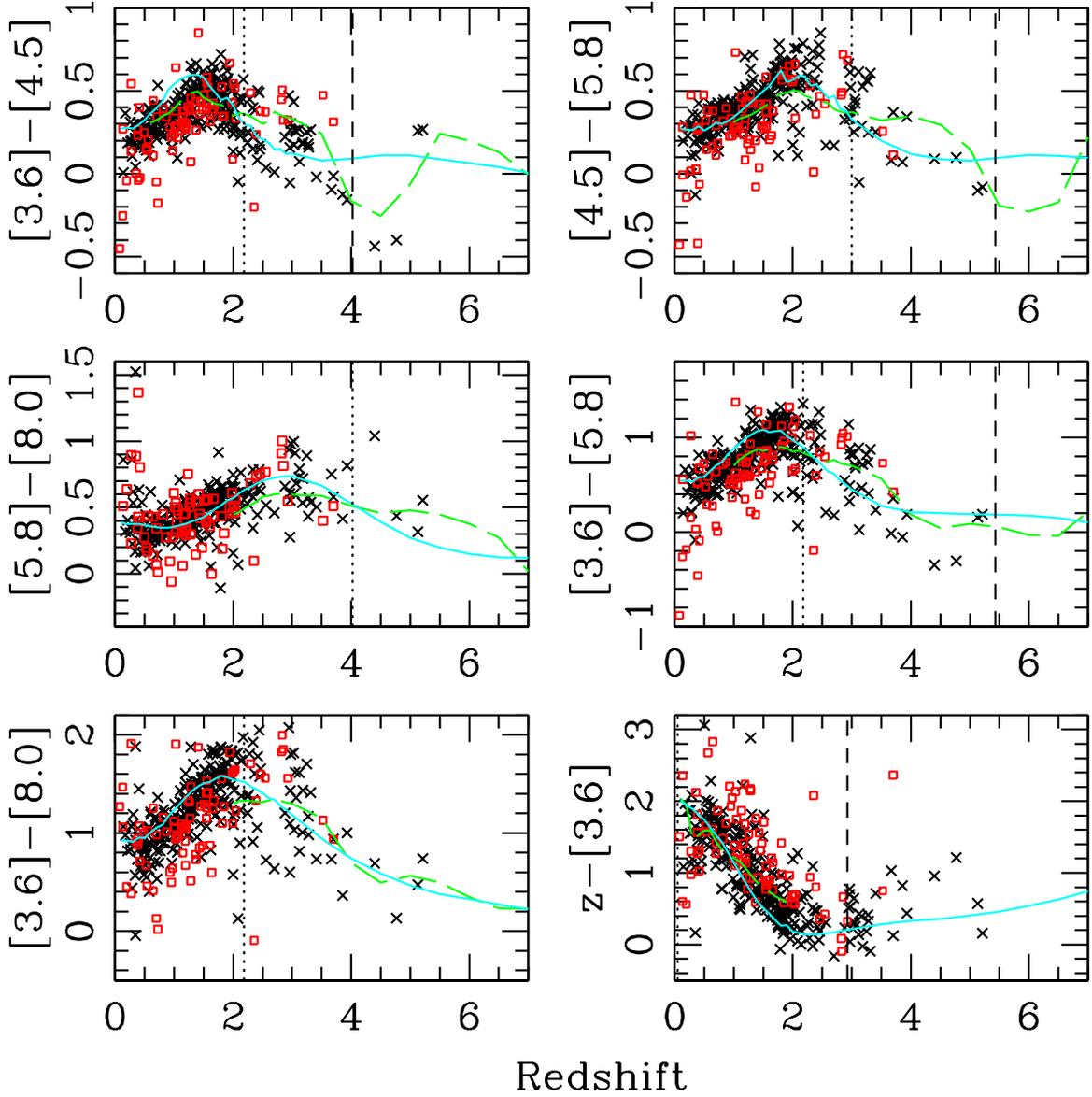}
\caption{Color-redshift relation for various combinations of Spitzer
colors.  Black symbols are SDSS-DR3 quasars from the XFLS and SWIRE
ELAIS-N1/N2 and Lockman Hole areas.  Red points are
Hectospec-confirmed XFLS quasars.  Errors are typically
$\sim0.14$\,mag.  The dashed green curve is the expected relation
found from convolving the geometric mean of the optical-IR composite
spectrum of \markcite{ghw05}{Glikman et al.} (2005) with the transmission curves.  The solid
cyan curve is the expected relation from convolving the \markcite{ewm+94}{Elvis} {et~al.} (1994)
composite radio-quiet SED with the transmission curves.  The vertical
lines mark the redshifts where 1$\mu$m enters the bluest ({\em
dotted}) and leaves the reddest ({\em dashed}) band.
\label{fig:fig3}}
\end{figure}

We begin by exploring the color-redshift relation for confirmed type 1
quasars.  Figure~\ref{fig:fig3} shows the color-redshift relation for
various combinations of Spitzer colors; see \markcite{rhv+03}{Richards} {et~al.} (2003, Fig.~2)
for similar plots using the SDSS bands.  In Figure~\ref{fig:fig3}
black points are SDSS-DR3 quasars from the XFLS and SWIRE ELAIS-N1/N2
and Lockman Hole areas.  Open red squares are Hectospec-confirmed XFLS
quasars.  The dashed green curve is the expected relation found from
convolving the geometric mean of the optical-IR composite spectrum of
\markcite{ghw05}{Glikman et al.} (2005) with the transmission curves.  The solid cyan curve is
the expected relation from convolving the \markcite{ewm+94}{Elvis} {et~al.} (1994) composite
radio-quiet SED with the transmission curves.  The {\em Spitzer}
colors are given in AB magnitude units where $[1]-[2] =
-2.5\log(S_1/S_2)$.

Much of the color change seen in Figure~\ref{fig:fig3} results from a
single strong feature in the typical quasar SED --- the so-called 1
micron inflexion, where the slope of the SED changes sign in a $\nu
f_{\nu}$ vs.\ $\nu$ representation.  Thus we have indicated (by dotted
and dashed lines, respectively) where the $1\mu$m inflexion enters
the bluest and leaves the reddest band.  We extend the plot well
beyond the redshifts of our confirmed quasars since {\em Spitzer}
should be sensitive to quasars with redshifts in excess of $z=7$, and
it is helpful to know the expected colors of such objects.

We show two sets of predicted color-redshift relations since, while
the \markcite{ewm+94}{Elvis} {et~al.} (1994) template covers the the full redshift range, it is
based in part on broad-band photometry, which smears out the effects
of the emission lines.  The spectroscopy-based curves derived from
\markcite{ghw05}{Glikman et al.} (2005) will be more accurate in regions of strong emission
lines, but this template covers a smaller wavelength (and thus
redshift) range.  For example the discrepancy at $z\sim3$ and
$z\sim4.5$ in the $[3.6]-[4.5]$ color results from the H$\alpha$ line
moving into and out of the filters.  Colors derived from non-adjacent
bandpasses show a much smaller effect as the emission lines are not
simultaneously moving out of the bluer band and into the redder band.
The smoothness of the color-redshift relations (as compared to the
SDSS) and the relatively large ($\sim10$\%) photometric errors and
means that IRAC colors alone may not be useful for accurate quasar
photometric redshift estimation \markcite{wrs+04}(e.g., {Weinstein} {et~al.} 2004), but MIR color
information in addition to optical colors will be extremely useful in
breaking redshift degeneracies in the optical.  Note in particular the
very large change in $z-[3.6]$ color between redshifts 0 and 2 as the
1$\mu$m inflexion moves between those two bandpasses.

\subsection{The Color-Color Relations}

\begin{figure}
\epsscale{1.0}
%\plotone{figures/iraccolors.ps}
\plotone{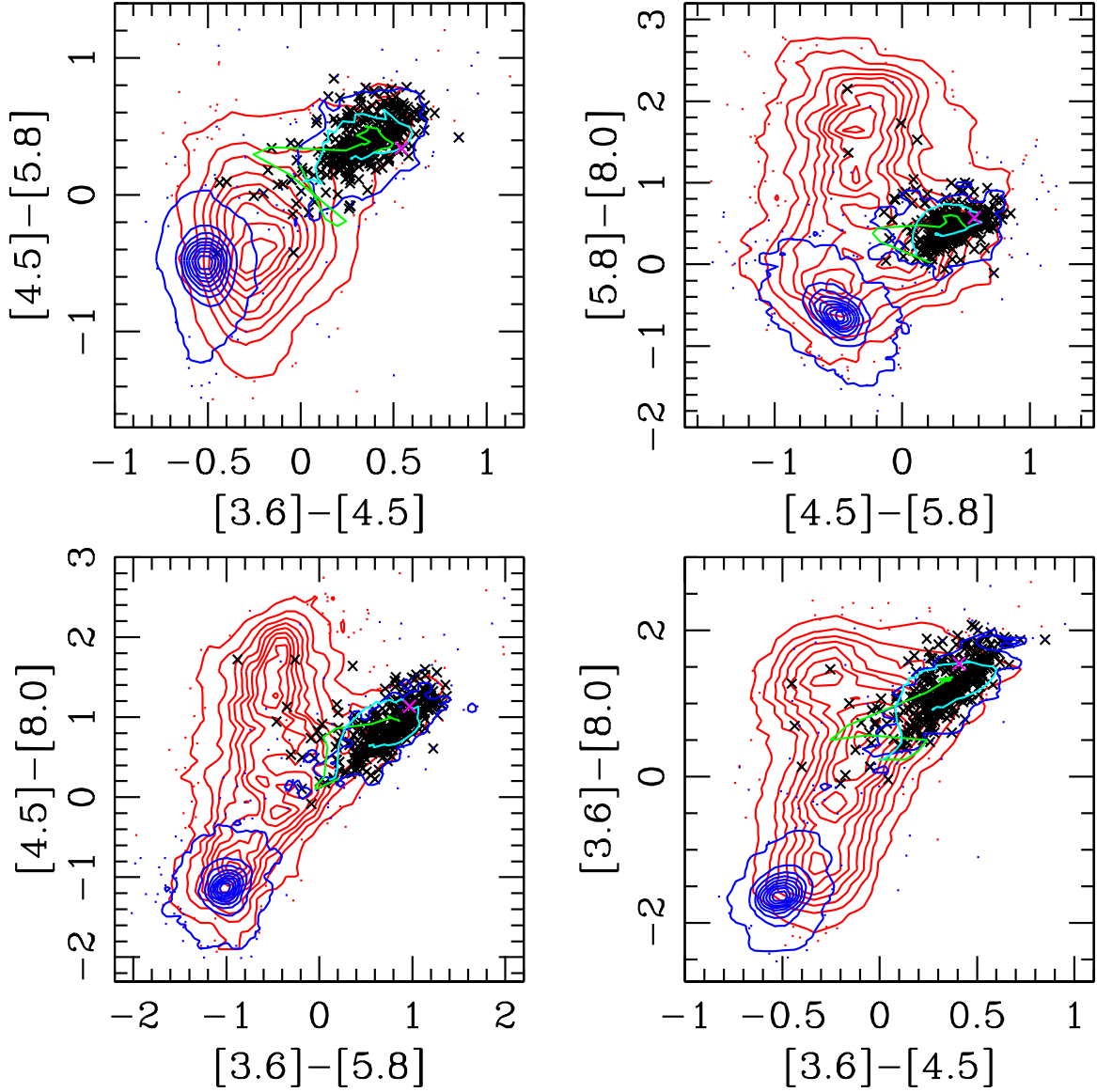}
\caption{IRAC colors of SDSS+{\em Spitzer} objects.  Blue contours and
dots correspond to objects that have point-like SDSS morphologies,
while objects classified by the SDSS as extended are shown by red
contours and dots.  These blue and red contours/dots represent all of
the objects detected in all four IRAC bands in the XFLS catalogs that
have a match in the SDSS-DR3 photometric database.  Black symbols
indicate spectroscopically confirmed type 1 quasars.  The few outliers
are from the Hectospec sample and may be mis-classified (or host
galaxy dominated).  The cyan lines are the $z=0.1$ to $z=7$ colors
predicted from the \markcite{ewm+94}{Elvis} {et~al.} (1994) quasar SED; the highest redshift
quasars are at the blue (negative) end of the tracks.  The green line
shows the predicted colors from $z=2$ ($z=1$ in the upper left-hand
panel) to $z=7$ using the \markcite{ghw05}{Glikman et al.} (2005) composite spectrum.  A magenta
cross indicates where the lowest redshift from the \markcite{ghw05}{Glikman et al.} (2005)
colors falls on the \markcite{ewm+94}{Elvis} {et~al.} (1994) curves.  Redder colors have more
positive values in this representation.
\label{fig:fig4}}
\end{figure}

We next examine the distribution of type 1 quasars in color-color
space.  \markcite{lss+04}{Lacy} {et~al.} (2004a), \markcite{seg+04}{Stern} {et~al.} (2005), and \markcite{hpp+04}{Hatziminaoglou} {et~al.} (2005) have shown
that AGN can be selected using only MIR colors such as those plotted
in Figure~\ref{fig:fig4}.  We plot all of the objects detected in all
four IRAC bands in the XFLS catalog that have a match in the SDSS-DR3
photometric database.  Blue contours and dots correspond to objects
that have point-like SDSS morphologies, while objects classified by
the SDSS as extended are shown by red contours and dots.  Black
symbols indicate the spectroscopically confirmed type 1 quasars.  The
cyan lines are the $z=0.1$ to $z=7$ colors predicted from the
\markcite{ewm+94}{Elvis} {et~al.} (1994) quasar SED; the highest redshift quasars have the
bluest (most negative) colors.  The green lines show the predicted
colors from the \markcite{ghw05}{Glikman et al.} (2005) composite spectrum from $z=2$ ($z=1$ in
the upper left-hand panel, which does not depend on the 8$\mu$m band)
to $z=7$.  The magenta cross indicates the redshift along the
\markcite{ewm+94}{Elvis} {et~al.} (1994) tracks where the \markcite{ghw05}{Glikman et al.} (2005) tracks start (either
$z=2$ or $z=1$).

\begin{figure}
\epsscale{1.0}
%\plotone{figures/mipscolors.ps}
\plotone{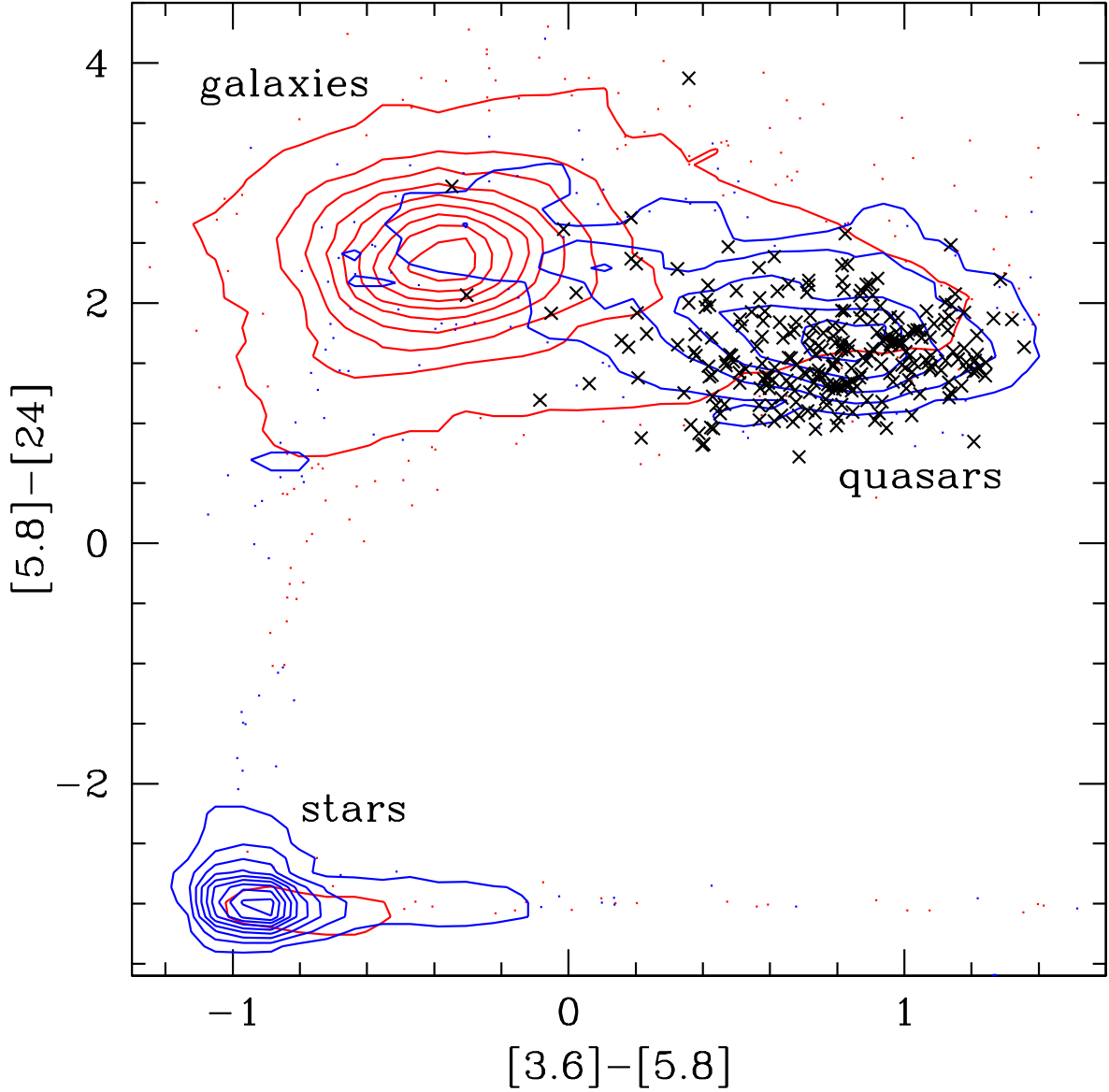}
\caption{$[3.6]-[5.8]$ vs.\ $[5.8]-[24]$ colors of SDSS+{\em Spitzer}
objects in the ELAIS areas.  Contours and points are as in
Fig.~\ref{fig:fig4}.  Stars (which have very blue $[5.8]-[24]$ color
and generally have point-like SDSS morphology classification) are much
more cleanly separated from quasars and galaxies when the MIPS
$24\mu$m data is included.  However, little is gained in terms of
galaxy-quasar separation as they have similarly red $[5.8]-[24]$
colors.
\label{fig:fig5}}
\end{figure}

From these plots, especially $[3.6]-[5.8]$ vs.\ $[4.5]-[8.0]$ in the
lower left-hand corner, it is clear that MIR colors combined with
optical morphology can select type 1 quasars efficiently (and with a
high degree of completeness) as point-like sources with red MIR
colors.  Luminous type 2 quasar candidates can also be identified as
those extended sources with AGN-like IR colors.  Adding the optical
morphology information confirms the speculation of \markcite{lss+04}{Lacy} {et~al.} (2004a) and
\markcite{seg+04}{Stern} {et~al.} (2005) regarding the nature of the two red branches, with
point-like AGN dominating the objects with red $[3.6]-[5.8]$ color and
extended sources (presumably low-redshift, PAH-dominated galaxies)
populating the red $[4.5]-[8.0]$ branch.  Adding this SDSS morphology
information thus enhances the MIR color-classification of sources
since the {\em Spitzer} PSF is too large to yield morphology
information for distant sources.  We further demonstrate the
classification information that is gained by including morphology in
Figure~\ref{fig:fig5} where we add MIPS $24\mu$m data in order to show
$[3.6]-[5.8]$ vs.\ $[5.8]-[24]$.  Adding the longer wavelength MIPS
information clearly helps to discriminate stars (which have very blue
$[5.8]-[24]$ colors) from galaxies and quasars which have very red
$[5.8]-[24]$ color); however, little is gained in terms of
quasar-galaxy separation that was not already realized with the IRAC
photometry.

While \markcite{hpp+04}{Hatziminaoglou} {et~al.} (2005) investigate both the MIR and optical colors of
their sample, the combined optical+MIR colors are not considered for
selection.  This is appropriate given that the flux density limits of
the {\em Spitzer} survey fields are much deeper than the spectroscopic
limit of SDSS quasar survey (for a typical quasar SED) and the IR is
much less affected by dust extinction (and thus is more representative
of the quasar population at a given nuclear magnitude).  However, the
SDSS photometry goes more than 3 magnitudes deeper than most of the
SDSS's spectroscopic survey for quasars, and at $i=21$, quasars must
be reddened by $E(B-V)\sim0.57$ to be extincted out of the SDSS
imaging in the $i$-band.  As such, it is interesting to explore the
selection of quasars using the combined SDSS plus {\em Spitzer} color
information.

\begin{figure}
\epsscale{1.0} 
%\plotone{figures/bothcolors34.eps}
\plotone{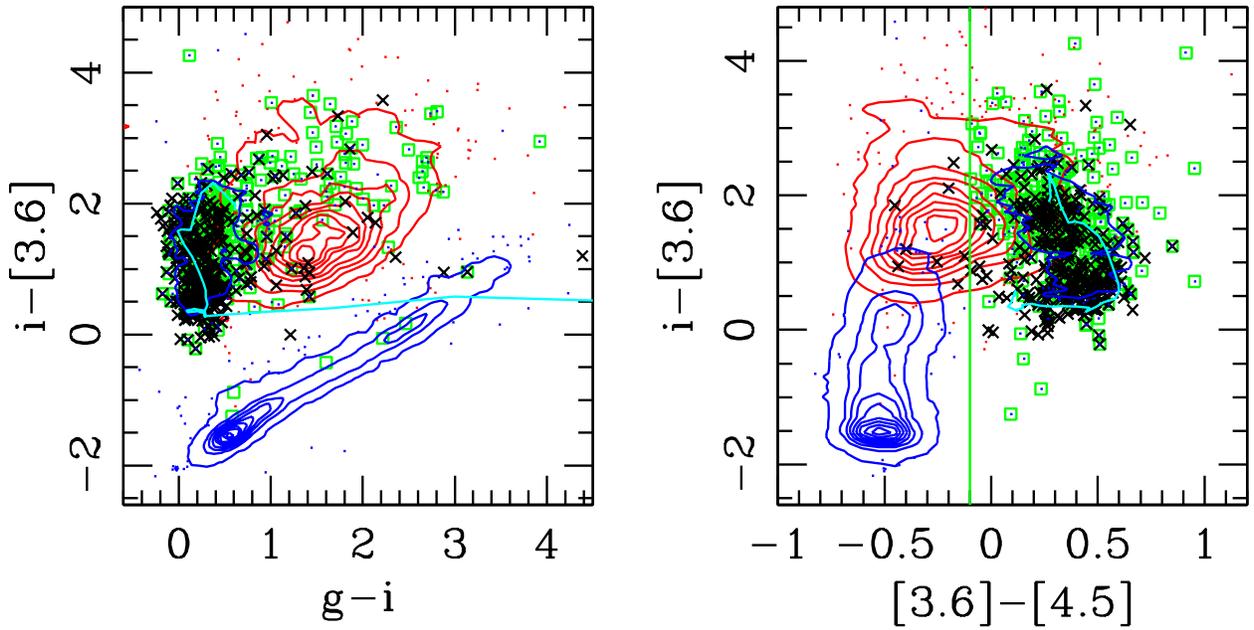}
\caption{Color-color matrix using an SDSS color ($g-i$), an SDSS+{\em
Spitzer} color ($i-[3.6]$), and a {\em Spitzer} color ($[3.6]-[4.5]$).
See Fig.~\ref{fig:fig4} for an explanation of the plot symbols.
Combining {\em Spitzer}-IRAC photometry with SDSS colors/morphology
defines a parameter space where stars, AGN, and normal galaxies can be
efficiently distinguished.  The green line in the right panel shows a
possible selection criterion for unresolved AGN; open green squares
show the point sources selected by such a cut.  Black symbols are
confirmed SDSS quasars. The cyan line shows the expected colors of
quasars from $z=0.1$ to $z=5$ (using the \markcite{ewm+94}{Elvis} {et~al.} 1994 mean SED);
low-redshift quasars having the reddest $i-[3.6]$ colors.  The
contours and dots are for XFLS sources only, whereas the black and
green symbols are drawn from all four areas of sky in our sample.
\label{fig:fig6}}
\end{figure}

Figure~\ref{fig:fig6} shows the color-space distribution of our
sources using an SDSS color, an SDSS+{\em Spitzer} color, and a {\em
Spitzer} color.  For the SDSS color, we choose $g-i$ since those are
the most widely separated high S/N bands in the SDSS filter set.  For
SDSS+{\em Spitzer}, we chose the two closest high S/N bands ($i$ and
$S_{3.6}$).  For the {\em Spitzer} color, we chose the two highest S/N
bands ($S_{3.6}$ and $S_{4.5}$); this choice happens to produce the
greatest separation of classes and has the added attraction that it
does not rely on the longer wavelength bands that will be lost when
{\em Spitzer's} coolant runs out.  This presentation shows that AGN,
normal galaxies, and stars can be separated quite cleanly when using
both optical and near-IR information, particularly when morphology
information is available in addition to the colors.  If we were simply
interested in selecting type 1 quasars, then selecting point sources
with $[3.6]-[4.5]>-0.1$ might be sufficient given that such a cut
includes nearly all of the point sources with non-stellar colors (with
the exception of quasars near $z=4.5$, see Fig.~\ref{fig:fig3}).
However, the morphology issue is complex.  Including morphology means
that normal galaxies will not contaminate the sample, but that type 2
AGN and faint AGN with poorly determined morphologies will be lost.
Judicious rotation of the axes in Figure~\ref{fig:fig6} may allow for
relatively clean AGN selection without having to rely on morphology
information.  However, a better approach would be to take advantage of
the 8-dimensional color space afforded by SDSS and {\em Spitzer}
photometry and performing a Bayesian classification such as the kernel
density estimation algorithm described by \markcite{rng+04}{Richards} {et~al.} (2004).  We intend
to pursue such an approach in a future publication.

\begin{figure}
\epsscale{1.0}
%\plotone{figures/bothcolors2.eps}
\plotone{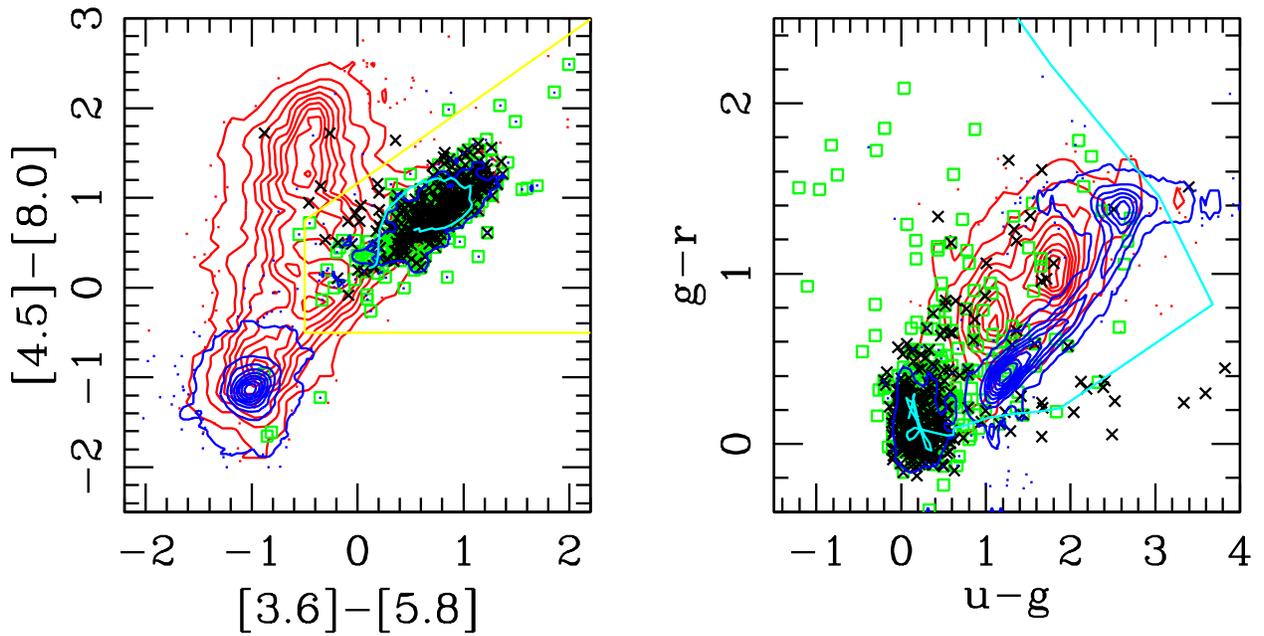}
\caption{MIR ({\em left}) and optical ({\em right}) color-color plots
showing the location of optical-color/morphology+MIR-color selected
populations.  See Fig.~\ref{fig:fig4} for an explanation of the plot
symbols.  Quasar candidates (see Fig.~\ref{fig:fig6}) are shown as
open green squares.  The yellow lines in the left-hand panel show the
MIR selection criteria from \markcite{lss+04}{Lacy} {et~al.} (2004a) [after having multiplied
their cuts by 2.5 to convert to a magnitude scale].
\label{fig:fig7}}
\end{figure}

We can compare this optical+MIR selection to the MIR-only selection of
\markcite{lss+04}{Lacy} {et~al.} (2004a), \markcite{seg+04}{Stern} {et~al.} (2005), and \markcite{hpp+04}{Hatziminaoglou} {et~al.} (2005).  Each of these
papers uses slightly different MIR selection techniques.  We will
compare explicitly only to \markcite{lss+04}{Lacy} {et~al.} (2004a) as their selection is seen
to empirically produce the greatest color separation with respect to
morphology.  In the left-hand panel of Figure~\ref{fig:fig7} we
reproduce the $[3.6]-[5.8]$ vs.\ $[4.5]-[8.0]$ color-color plot from
\markcite{lss+04}{Lacy} {et~al.} (2004a), showing their selection as yellow lines.  Objects
selected as being optical point sources with red MIR colors are shown
as open green squares.  Adding the optical data to the selection
criteria allows us to select quasars more efficiently (with reduced
completeness due to the loss of AGN dominated by host galaxy light in
the optical).  The right-hand panel of Figure~\ref{fig:fig7} shows
where these objects lie in $u-g$ vs.\ $g-r$ color space.  For the sake
of clarity, we have limited the point and extended sources to $g<21$
to reduce the scatter due to SDSS photometric error, but the green
squares (point-like quasar candidates) have no such limit applied.  We
note that, despite the MIR color selection, the green squares are
still predominantly UV-excess sources.  To $i<21$, we find that 70\%
of the MIR color-selected quasars have $u-g<0.6$ and $g-r<0.6$.  Thus,
the fraction of dust-reddened/extincted type 1 quasars
\markcite{rhv+03,ggl+04}(e.g., {Richards} {et~al.} 2003; {Glikman} {et~al.} 2004) that might be missed by an optical
survey with UVX color-selection is no larger than 30\%.  This limit
applies to quasars with $i$-band extinction less than 1.0 mag (the
difference between the SDSS $i$-band imaging limit and our adopted cut
of $i=21$), which is $E(B-V)=0.48$ for Galactic extinction.  This
fraction may be higher if there are type 1 quasars with larger
extinction, but is most likely smaller than 30\% since many of the
non-UVX quasars will simply have $z\gtrsim2.2$.  Quasars with $z>2.2$
have redder optical colors even if they are not dust-reddened; this
population will still be identified by the SDSS quasar-selection
algorithm.

\begin{figure}
\epsscale{1.0}
%\plotone{figures/bothcolormag2.eps}
\plotone{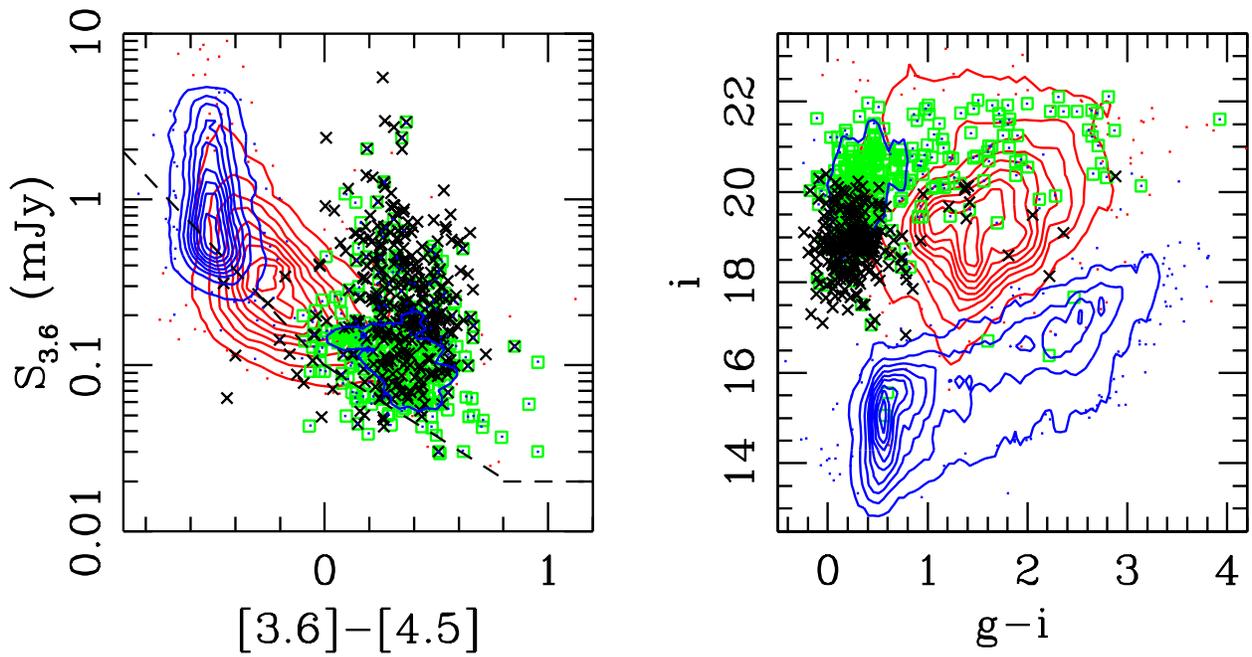}
\caption{{\em Left:} {\em Spitzer} color-magnitude diagram.  The
dashed lines show the effects of the 5.8$\mu$m and 8.0$\mu$m flux
density limits.  {\em Right:} SDSS color-magnitude diagram.  Faint
($i\gtrsim18$) point sources with $g-i\gtrsim0.8$ are either
high-redshift quasars or dust-reddened low-redshift quasars. In both
panels, the contours/points are coded the same as in
Fig.~\ref{fig:fig6}.
\label{fig:fig8}}
\end{figure}

Figure~\ref{fig:fig8} shows color-magnitude plots for this selection
criterion as compared with known quasars and point/extended sources in
general.  The 3.6$\mu$m flux density is limited for $[3.6]-[4.5] <
0.0$ by the 8.0$\mu$m flux density limit and for $0.0 < [3.6]-[4.5] <
0.8$ by the 5.8$\mu$m flux density limit, as shown by the dashed
lines.  Note that while the 3.6$\mu$m flux density limit is 20$\mu$Jy,
a 3.6$\mu$m selected sample is complete to all AGNs to only
$\sim200\mu$Jy.  In the right-hand panel we again see that the density
of dust-reddened and extincted type 1 quasars cannot be too large
since the density of quasars with $i<19$ and $g-i<0.8$ is much higher
than that of $i<22$ quasar candidates with $g-i>0.8$.  This figure
also serves to illustrate that MIR-only selection is most efficient
for bright MIR sources.  For $S_{3.6}>1$mJy there is little
contamination of the MIR color space of AGNs, but at fainter limits,
contamination from star-forming galaxies becomes problematic without
additional selection information (such as the optical morphology).

\section{The Obscured Quasar Fraction}

\begin{figure}
\epsscale{1.0}
%\plotone{figures/limitssed.eps}
\plotone{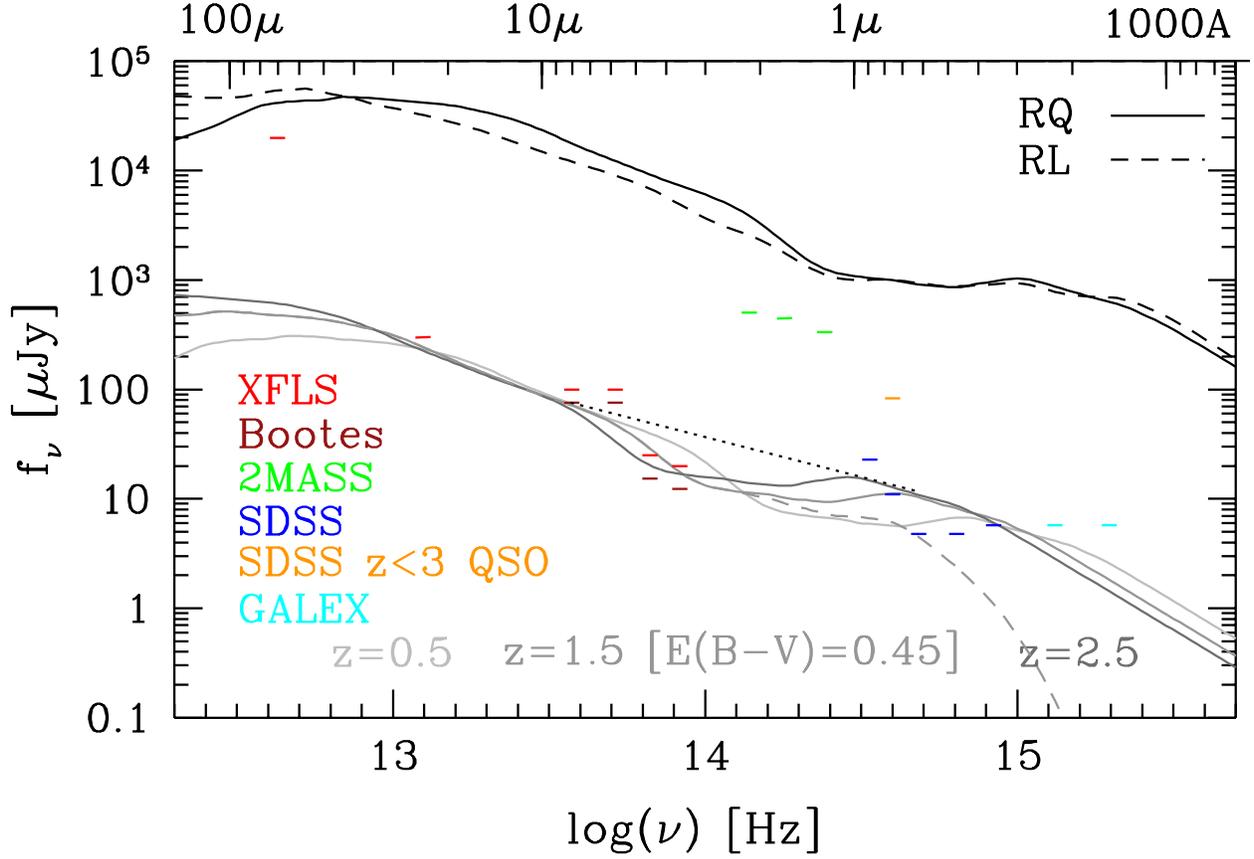}
\caption{Mean quasar SEDs compared to various survey flux limits.  The
black solid and dashed lines are the $z=0$ \markcite{ewm+94}{Elvis} {et~al.} (1994) mean
radio-quiet and radio-loud SEDs, normalized to $i=16.4$ (1 mJy).  The
colored dashes show the flux density limits for the {\em
Spitzer}-XFLS, {\em Spitzer}-Bo\"{o}tes \markcite{seg+04}({Stern} {et~al.} 2005), 2MASS, SDSS
(imaging), SDSS (low-z quasar spectroscopy) and {\em GALEX}.  The
lighter gray lines show the radio-quiet SED at $z=0.5$, $1.5$, and
$2.5$ normalized to the Bo\"{o}tes field $8\mu$m flux limit (dashed
gray line is $z=1.5$ reddened by $E(B-V)=0.25$).  The dotted line
shows a spectral index of $\alpha_{\nu}=-0.73$ between $8\mu$m and the
SDSS $r$ band.
\label{fig:fig9}}
\end{figure}

Here we make some comparisons between mid-IR only selection and
optical+mid-IR selection in terms of relative flux limits.
Figure~\ref{fig:fig9} shows mean quasar SEDs compared to various
survey flux limits.  The black solid and dashed lines are the
(rest-frame) \markcite{ewm+94}{Elvis} {et~al.} (1994) mean radio-quiet and radio-loud SEDs,
normalized to $i=16.4$ (1 mJy).  The colored dashes show the flux
density limits for the {\em Spitzer}-XFLS, {\em
Spitzer}-NOAO/Bo\"{o}tes \markcite{jd99,esb+04}({Jannuzi} \& {Dey} 1999; {Eisenhardt} {et~al.} 2004), 2MASS, SDSS (imaging),
SDSS ($z<3$ quasar spectroscopy), and {\em GALEX}.  The lighter gray
lines show the radio-quiet SED at z=$0.5$, $1.5$, and $2.5$ normalized
to the Bo\"{o}tes field $8\mu$m flux limit.  The dashed gray line
shows the $z=1.5$ SED reddened by SMC-like reddening with
$E(B-V)=0.45$.  The {\em Spitzer} flux density limits of the
Bo\"{o}tes and FLS areas are shown as brown and red bars.  The 2MASS
limits are given in green.  The SDSS imaging limits are shown in blue,
with the spectroscopic limit shown in orange.  Cyan bars show the {\em
GALEX} flux density limits (for the Medium Imaging Survey [MIS]).

\markcite{seg+04}{Stern} {et~al.} (2005) estimate the fraction of obscured quasars missed by
optical surveys by comparing their $8\mu$m source counts to the source
counts of quasars at $R\sim21$, assuming a spectral index of
$\alpha_{\nu}=-0.73$.  We revisit this analysis with two new
considerations.  First, we use the number counts of optically selected
($g$-band) quasars from \markcite{rca+05}{Richards} {et~al.} (2005) which used SDSS imaging coupled
with 2dF spectroscopy to determine the quasar luminosity function to
$g\sim21.5$.  Second, rather than assuming a power-law SED, we use an
empirical SED such as that from \markcite{ewm+94}{Elvis} {et~al.} (1994).

\markcite{seg+04}{Stern} {et~al.} (2005) reported an obscured quasar fraction of $(275-65)/275 =
76$\%, where 275 is their candidate AGN density per square degree and
65 is the density of optically selected AGN to $R=21$ from
\markcite{wwb+03}{Wolf} {et~al.} (2003).  We find that a typical quasar with an 8$\mu$m flux
density of 76$\mu$Jy has a $g$-band magnitude of 21.5.  Using the
quasar luminosity function of \markcite{rca+05}{Richards} {et~al.} (2005), we find an optically
selected quasar density of 89 per square degree to $g=21.5$.  This
density gives an upper limit to the obscured fraction of $(275-89)/275
= 68$\%.

However, since the quasar SED is {\em not} a power law between the
optical and MIR, but rather exhibits some significant features (such
as the 1$\mu$m inflexion), it is important to determine the $g$-band
flux limit as a function of redshift.\footnote{It is probably equally
as important to consider the obscured quasar fraction as a function of
luminosity \markcite{uao+03,hsf+05}({Ueda} {et~al.} 2003; {Hao} {et~al.} 2005), but that is beyond the scope of our
analysis.}  The expected flux density of a type 1 quasar with an
8$\mu$m flux density of 76$\mu$Jy can reach as faint as $g=22.5$
depending on redshift (and thus the optical counts will be
underestimated).  At that magnitude limit there are 170 optically
selected quasars per square degree.  We illustrate this issue in
Figure~\ref{fig:fig9} where one can see that at $z=0.5$ the expected
$g$ and $r$ flux densities of a typical quasar are much fainter (but
still detectable by SDSS) than the optical-to-MIR flux ratio assumed
by \markcite{seg+04}{Stern} {et~al.} (2005) (dotted line).  While the density of luminous
quasars peaks at $z\sim2.5$ where the \markcite{seg+04}{Stern} {et~al.} (2005) extrapolation is
roughly correct, X-ray surveys \markcite{uao+03}(e.g., {Ueda} {et~al.} 2003) have shown that
the density of lower luminosity AGNs peaks at lower redshifts and thus
this effect is important.

If we instead use our the mean SED that we will derive in
\S~\ref{sec:meansed}, the $g$-band magnitude for a 76$\mu$Jy source at
$8\mu$m ranges between $g=21.4$ and $g=22.4$.  The optically selected
quasar densities for those magnitudes are 82 and 161 per square
degree, respectively.  Using these numbers, we find that the fraction
of obscured quasars has a lower limit in the ballpark of
$(275-161)/275 = 41$\%.  Finally, since the \markcite{seg+04}{Stern} {et~al.} (2005)
MIR-selected AGNs were selected to a rather faint MIR flux limit,
then, as discussed above, some fraction of the AGN candidates may
instead turn out to be normal star forming galaxies, which would also
reduced the obscured quasar fraction.

Similar arguments can be made with regard to the obscured quasar
fraction computed by \markcite{lss+04}{Lacy} {et~al.} (2004a).  They found an obscured fraction
of $(16/35)=46$\% by assuming that all 8$\mu$m sources brighter than
1\,mJy would be brighter than the SDSS's spectroscopic magnitude limit
of $i=19.1$.  Using the \markcite{ewm+94}{Elvis} {et~al.} (1994) SED to determine the $i$
magnitude for this 8$\mu$m flux density as a function of redshift, we
find that many IR-bright, low redshift quasars will in fact be fainter
than the SDSS spectroscopic limit.  Again, this effect is largest at
low redshift, but \markcite{lss+04}{Lacy} {et~al.} (2004a) find their obscured-AGN candidates are
primarily low redshift.  
%Using the redshifts given in Table~1 of
%\markcite{lss+04}{Lacy} {et~al.} (2004a), we find that as many as 4 of the 16 obscured AGNs
%candidates may be fainter than the SDSS spectroscopic magnitude limit
%when using the \markcite{ewm+94}{Elvis} {et~al.} (1994) SED to determine the optical flux limit
%from the 8$\mu$m flux limit.  This number increases from 4 to 8 for
%our new mean SED.  
In principle, this would suggest that the obscured quasar fraction
found by \markcite{lss+04}{Lacy} {et~al.} (2004a) is also an upper limit; however, in practice, their
objects do appear to be bona-fide obscured AGNs \markcite{lcr+04}({Lacy et al.} 2004b).  This
may be due in part to the much brighter MIR flux limit imposed by
\markcite{lss+04}{Lacy} {et~al.} (2004a) (as compared to \markcite{seg+04}{Stern} {et~al.} 2005).  Nevertheless, we
caution that the shape of the SED can influence the expected
MIR-to-optical ratio of quasars as a function of redshift.

Other work on obscured AGN using Spitzer selection has typically
suggested a large obscured quasar fraction. \markcite{mrl+05}{Mart{\'{\i}}nez-Sansigre} {et~al.} (2005) use a
combined radio and {\em Spitzer} 24$\mu$m based selection in the XFLS
to show that the ratio of obscured to type-1 quasars is in the range
1--$3:1$, but this estimate is dependent on a lack of evolution in the
distribution of the ratio of radio to optical luminosities of quasars
from low to high redshifts, as the majority of high redshift quasars
are expected to have much lower radio luminosities than probed by the
radio survey of the XFLS by \markcite{ccy+03}{Condon} {et~al.} (2003).  The deep,
multi-wavelength Great Observatories Origins Deep Survey (GOODS) finds
a ratio of obscured to unobscured AGN of $\sim 3:1$ \markcite{tuc+04}({Treister} {et~al.} 2004),
but most of their AGN are well below quasar luminosity, and thus not
directly comparable to the objects studied in this paper.  It is also
important to realize that mid-infrared selection may not find all
obscured quasars, particularly at high redshift, where the IRAC bands
are redshifted into the rest-frame near infrared, or in cases of
objects viewed through very high ($A_V\sim 100$) extinctions, as
expected for objects viewed through an obscuring torus edge-on to the
line of sight. Improved modeling of the biases in mid-infrared
selection of AGN should remove much of this uncertainty.

\section{Spectral Energy Distributions}
\label{sec:sed}

\subsection{Introduction}

Accurate determination of quasar SEDs is crucial to our understanding
of quasar physics.  The strengths and shapes of the various components
of a typical quasar SED have been used to determine the physical
drivers in each frequency regime from the jets in the radio to the
dusty torus in the mid-to far-IR, the accretion disk in the
optical/UV/soft-X-ray, and the X-ray corona at the highest energies.

As the observed radiation from quasars is often re-processed and
re-emitted at wavelengths far different from the seed photons, knowing
the overall SED is necessary for determining bolometric luminosities.
In turn, bolometric luminosities are necessary for determining the
so-called Eddington ratio ($L_{\rm bol}/L_{\rm Edd}$, or equivalently
$L_{\rm bol}/M_{\rm BH}$), which serves as a measure of the accretion rate of
the system \markcite{pet97}(e.g., {Peterson} 1997).

Despite a few recent additions/updates to \markcite{ewm+94}{Elvis} {et~al.} (1994),
\markcite{pch+00,kwh+03,re04}(e.g., {Polletta} {et~al.} 2000; {Kuraszkiewicz} {et~al.} 2003; {Risaliti} \& {Elvis} 2004), complete SEDs have been compiled
for only a small number ($\lesssim100$) of quasars and the mean SED
from \markcite{ewm+94}{Elvis} {et~al.} (1994) is arguably still the best description of the SED
of quasars and is certainly the most commonly used.  However, it is
illuminating to recall the caveats given by the authors in the
abstract of that paper: ``We derive the mean energy distribution (MED)
for radio-loud and radio-quiet objects and present the bolometric
corrections derived from it.  We note, however, that the dispersion
about this mean is large ($\sim1$ decade for both the infrared and
ultraviolet components when the MED is normalized at the near-infrared
inflection). ... The existence of such a large dispersion indicates
that the MED reflects on some of the properties of quasars and so
should be used only with caution.''

Furthermore, in addition to the SED diversity in their sample, it is
well known that the sample used to construct this mean SED is not
entirely representative: ``Our primary selection criteria were (1)
existing {\em Einstein} observations at good signal-to-noise ratio (to
ensure good X-ray spectra) and (2) an optical magnitude bright enough
to make an {\em IUE} spectrum obtainable.''  The first criterion means
that the sample is biased toward towards X-ray bright quasars, and
therefore typically objects with relatively high X-ray-to-optical flux
ratios.  Similarly, the second criterion biases the sample towards
bluer quasars \markcite{jsr+05}(see {Jester} {et~al.} 2005).

Nevertheless, the mean SED from \markcite{ewm+94}{Elvis} {et~al.} (1994) has generally been
found to be a robust template for typical type 1 quasars.  However,
given the diversity of the mean SED, the biases inherent to the
sample, and the accuracy to which we would like to know the bolometric
luminosity for a given quasar, it is important to improve our
knowledge of the full SED of quasars to test how well this assumption
holds and to investigate the full range of individual SEDs.

\begin{figure}[p]
\epsscale{1.0}
%\plotone{figures/sedtheoryboth.ps}
\plotone{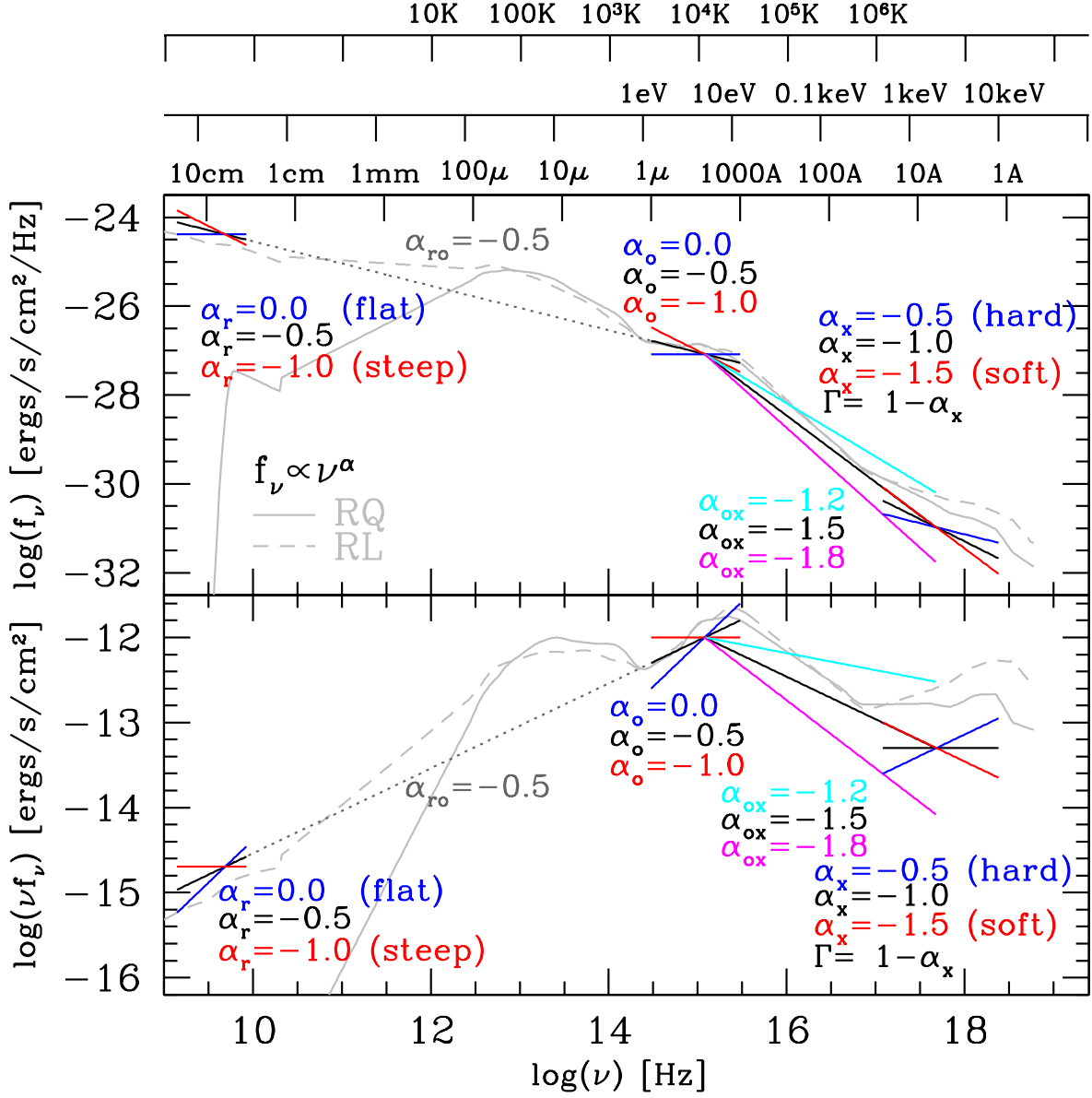}
\caption{Quasar SED diagnostic plot.  Show in gray are the
\markcite{ewm+94}{Elvis} {et~al.} (1994) radio-quiet (solid) and radio-loud (dashed) mean SEDs.
The colored lines indicate typical spectral indices in the radio,
optical, and X-ray using the same sign convention.  Also shown is the
typical radio-to-optical spectral index for radio-loud quasars and the
range of optical-to-X-ray spectral indices. Studies in different bands
tend to use different sign conventions for spectral indices and jargon
to describe them (e.g., steep/red/soft).  The top panel shows
$f_{\nu}$, while the bottom panel shows $\nu f_{\nu}$.  The $x$-axis is
labeled as log($\nu$) (bottom), and wavelength, energy, and temperature
(top).
\label{fig:fig10}}
\end{figure}

As there is often little overlap in the jargon and notation used to
describe quasars in different bandpasses
(e.g. blue/red/Angstroms/magnitudes in the optical, flat/steep/GHz/mJy
in the radio, hard/soft/keV/cgs in the X-ray), it will aid our
discussion to review quasar SEDs from a single multi-wavelength point
of view.  Thus in Figure~\ref{fig:fig10} we show the mean
radio-loud and radio-quiet quasar SEDs from \markcite{ewm+94}{Elvis} {et~al.} (1994) on a
log-log plot of $f_{\nu}$ vs.\ ${\nu}$ (top panel) and $\nu f_{\nu}$
vs.\ ${\nu}$ (bottom panel).  Here $f_{\nu}$ is given in
ergs\,s$^{-1}$\,cm$^{-2}$\,Hz$^{-1}$, and the abscissa is given in log
frequency, wavelength, energy, and temperature units.  The
normalization is such that the quasar has a continuum flux density of
0.08\,mJy ($AB=19.1$) at 2500\AA.  Typical ranges of spectral indices
within and between bands are shown; the spectral indices are given as
energy indices where $f_{\nu} \propto \nu^{\alpha}$ such that $\alpha$
represents the {\em slope} (and is thus generally negative).

We wish to use Figure~\ref{fig:fig10} to make a few points.
First, there is considerable scatter in quasar SEDs in each energy
regime and little is known about correlations in these variations {\em
between} energy regimes.  Second, the plots visually emphasize the
point made above, which is that the \markcite{ewm+94}{Elvis} {et~al.} (1994) sample was quite
X-ray bright.  Finally, there is a strong need for better
characterization of quasar SEDs in the near- through far-IR.

\subsection{Mean SEDs}
\label{sec:meansed}

To explore the color and luminosity dependence of the quasar SED, we
use the 259 SDSS quasars cataloged herein to construct new mean SEDs.
The individual SEDs of these quasars are shown in the Appendix.  For
the sake of homogeneity, we have excluded the Hectospec-confirmed
quasars from \markcite{pap+05}{Papovich et al.} (2005) as they were selected with different
criteria than the SDSS-confirmed quasars.

These SEDs are constructed as follows.  While all of our objects have
5 SDSS magnitudes and 4 {\em Spitzer}-IRAC flux densities, many
objects have no measurements in the 10 other bands that we catalog.
Thus we first attempt a form of ``gap-repair''.  For the {\em GALEX}
$f$- and $n$-bands, $J$, $H$, and $K$, the {\em ISO} 15$\mu$m-band,
and the {\em Spitzer}-MIPS 24 and 70$\mu$m bands, we replace missing
values with those determined from normalizing the \markcite{ewm+94}{Elvis} {et~al.} (1994) SED in
the next nearest bandpass for which we have data.  For example, $n$ is
estimated from $u$, $J$ from $z$, $K$ from $S_{3.6}$, $S_{15}$ from
$S_{8.0}$, etc.  For the radio-loud quasars we use the \markcite{ewm+94}{Elvis} {et~al.} (1994)
mean radio-loud SED; the remaining quasars are repaired using the
\markcite{ewm+94}{Elvis} {et~al.} (1994) mean radio-quiet SED.

Since the optical-to-X-ray flux ratio of the \markcite{ewm+94}{Elvis} {et~al.} (1994) SED is
somewhat abnormal, we do not perform the same sort of gap repair in
the X-ray for those quasars that lack {\em ROSAT} detections. Instead,
we use the $L_{UV}$--$L_X$ relationship that has been known since
\markcite{at86}{Avni} \& {Tananbaum} (1986) and has been well-characterized by \markcite{sbs+05}{Strateva} {et~al.} (2005).  This
relationship parameterizes the relatively tight correlation between
the 2500\AA\ and 2\,keV brightness of quasars as a function of
luminosity.  Thus we normalize the \markcite{ewm+94}{Elvis} {et~al.} (1994) SED to the average of
the high S/N ($gri$) SDSS measurements and determine the 2500\AA\ flux
density for that normalization.  Then we use the equations given by
\markcite{sbs+05}{Strateva} {et~al.} (2005) to estimate the 2\,keV flux density of each quasars.
We then assume a flat (in $\nu f_{\nu}$ space, which has
$\alpha_x=-1.0$) spectrum between 0.1 and 10\,keV, roughly consistent
with \markcite{rt00}{Reeves} \& {Turner} (2000) and \markcite{gty+00}{George} {et~al.} (2000).  Finally, as there are only 8
radio-loud quasars in our sample and wavelengths shortward of
$\sim100\mu$m are energetically unimportant, we ignore any missing
radio information (though all the objects have upper limits).

In addition to gap-repair, it is necessary to correct for host galaxy
contamination, which can have a significant effect in many of the
bandpasses that we consider.  Lacking the data to measure the host
galaxies of our sources, we must estimate their contribution.  Such
estimates of the contribution of host galaxy light to the SEDs can be
made by applying simple scaling relationships among host bulge
luminosity, bulge mass, black hole mass, and Eddington luminosity
\markcite{dmk+03,vsy+06}(e.g., {Dunlop} {et~al.} 2003; {Vanden Berk} {et~al.} 2006).  The well-known correlation between
central black hole mass and both bulge mass and luminosity
\markcite{fer02}(e.g., {Ferrarese} 2002) implies that quasars accreting at near their
Eddington limits must have host bulge masses, and therefore
luminosities, large enough to harbor the inferred black hole.  We have
used the quasar luminosity vs.\ host luminosity relationship at optical
frequencies, described by \markcite{vsy+06}{Vanden Berk} {et~al.} (2006), to estimate the relative
host galaxy contribution, assuming that the quasars are emitting at
their Eddington limits.

We specifically convert the $M_r$ relationship used by \markcite{vsy+06}{Vanden Berk} {et~al.} (2006)
to a luminosity and determine the host galaxy luminosity as
\begin{equation}
\log(L_{\rm Host}) = 0.87\log(L_{\rm AGN}) + 2.887 - \log(L_{\rm Bol}/L_{\rm Edd}),
\end{equation}
where the host galaxy and AGN luminosities are specific luminosities
at 6156\AA\ (the effective wavelength of the SDSS $r$ band) and we
have taken $L_{\rm Bol}/L_{\rm Edd}$ to be unity.  This synthesis of
multiple scaling relations will have considerable scatter, and many,
if not most, AGNs will be accreting at much lower Eddington ratios.
However, we have found that a value of $(L_{\rm Bol}/L_{\rm Edd})=1$
works well and it has the nice feature that it represents the {\em
minimum} host galaxy contribution that {\em must} be accounted for
(smaller ratios implying relatively more luminous hosts).

This process sets the relative scaling of the host galaxy in the
optical bandpasses.  To actually subtract the host galaxy contribution
at all wavelengths, we use the elliptical galaxy template of
\markcite{fr97}{Fioc} \& {Rocca-Volmerange} (1997) scaled according to the prescription above.  We ignore
the differences between spiral and elliptical hosts as the host galaxy
contribution is small where these differences matter most.  Using this
template and the above scaling, we find that at its peak
($\sim1.6\mu$m), the host galaxy contributes between 30 and 38\% of
the total observed luminosity.  This fraction approaches unity for
$(L_{\rm Bol}/L_{\rm Edd})=1/3$.  In practice, we have removed the
host galaxy contribution before applying the above gap-repair process
since the \markcite{ewm+94}{Elvis} {et~al.} (1994) template has already been corrected for the
host galaxy contribution.

As we wish to investigate the mean properties of quasars as a function
of luminosity, color, etc., we next combine the corrected individual
SEDs and construct a number of different mean SEDs.  We first convert
the flux densities of each individual object to luminosities in our
adopted cosmology and shift the bandpasses to the rest-frame.  We then
create a grid of points separated by 0.02 in log frequency and
linearly interpolate between the 19 rest-frame effective frequencies
that are populated by our gap-repaired photometry.

\begin{figure}[p]
\epsscale{1.0}
%\plotone{figures/sedplot.eps}
\plotone{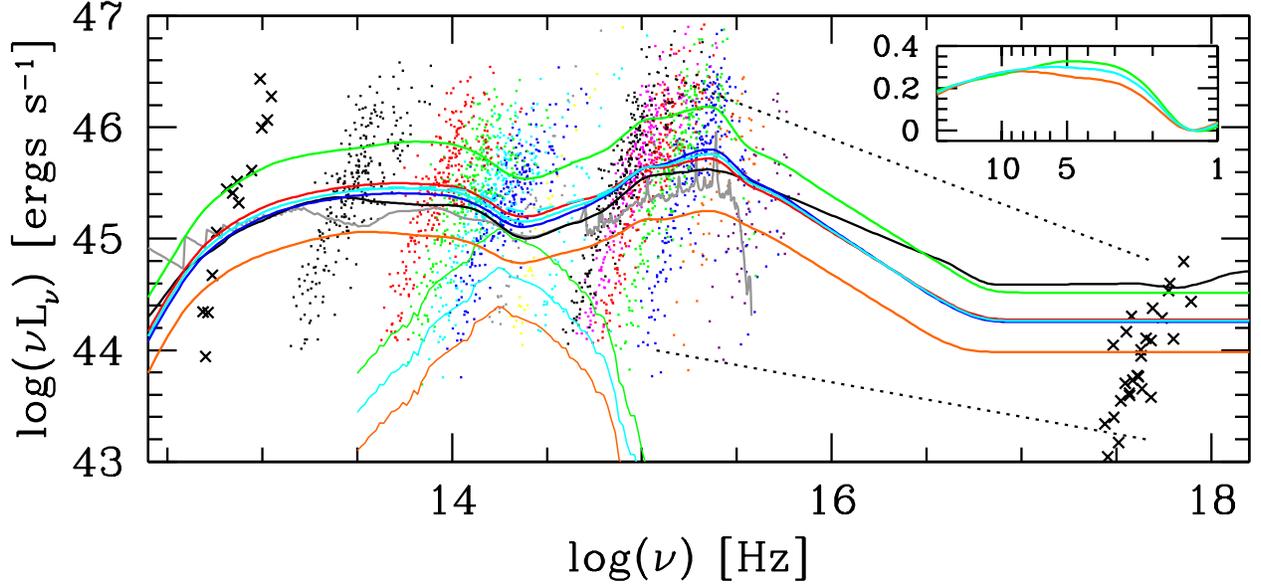}
\caption{Mean quasar SEDs and the data used to construct them.  The
points show the data; the lines show the mean SEDs.  From left to
right: black crosses--MIPS70, black points--MIPS24/ISO15, red
points--$S_{8.0}$, green points--$S_{5.8}$, cyan points--$S_{4.5}$,
blue points--$S_{3.6}$, gray points--$K$, yellow points--$H$, cyan
points--$J$, black points--$z$, magenta points--$i$, red points--$r$,
green points--$g$, blue points--$u$, orange points--$n$, purple
points--$f$, black crosses--x-ray.  The dotted lines show the range of
$\alpha_{\rm ox}$ with luminosity for the luminosity extremes of our
sample \markcite{sbs+05}({Strateva} {et~al.} 2005).  From top to bottom (at $10^{14}$ Hz) the thick
solid lines are green--optically luminous SDSS quasars, red--optically
red SDSS quasars, cyan--all SDSS quasars, blue--optically blue SDSS
quasars, black--\markcite{ewm+94}{Elvis} {et~al.} (1994) radio-quiet mean SED,
gray--\markcite{hpp+04}{Hatziminaoglou} {et~al.} (2005) mean SED (normalized to \markcite{ewm+94}{Elvis} {et~al.} 1994 at
$1\mu$m), orange--optically dim SDSS quasars.  The near-IR
luminous/dim composites are nearly identical to the optical
composites.  The thin green, cyan and orange curves show the host
galaxy contribution assuming $(L_{\rm Bol}/L_{\rm Edd})=1$ (see Eq.~1)
for the elliptical galaxy composite spectrum of \markcite{fr97}{Fioc} \& {Rocca-Volmerange} (1997).  The
dashed cyan curve shows the effect of ignoring the host contribution.
The inset at the upper right zooms in on the MIR region of the
spectrum.  The three curves are normalized at 1.3$\mu$m and the
$y$-axis shows relative luminosity, while the $x$-axis is in microns.
\label{fig:fig11}}
\end{figure}

The resulting mean SEDs are shown in Figure~\ref{fig:fig11}, where the
mean radio-quiet SED from \markcite{ewm+94}{Elvis} {et~al.} (1994) is shown in black and the
mean SED from \markcite{hpp+04}{Hatziminaoglou} {et~al.} (2005) is shown in gray (normalized to the black
curve at $1\mu$m).  The individual data points for our quasars are
also shown, color-coded by band.  The mean of all our objects is shown
in cyan.  We have further constructed mean SEDs for the optically
luminous/dim and optically blue/red halves of the population as shown
in green, orange, blue, and red, respectively.  The median values used
to split the distributions are optical: $\log(L)=46.02$; IR:
$\log(L)=46.04$; color: $\Delta(g-i)=-0.04$; where extremely red
quasars having $\Delta(g-i)>0.3$ were excluded from the construction
of all the composite SEDs (19 quasars in all).  These mean SEDs are
given in tabular form in Table~\ref{tab:tab3}.  The thin lines show
the relative host galaxy contribution for the luminous (green), all
(cyan), and dim (orange) samples using the scaling and galaxy template
as described above.  The dashed cyan line shows the effect of ignoring
the host contribution for the full-sample mean SED.

One of the primary purposes of constructing these mean SEDs as a
function of various quasar properties is to see how the MIR part of
the SED changes with these properties.  Interestingly, we find that
the {\em shape} of the MIR part of the SED is very similar for
optically blue and optically red quasars.  However, there are
significant differences between the most- and least-optically luminous
quasars in our sample.  The inset of Figure~\ref{fig:fig11} shows
that the luminous quasars are much brighter in the 4$\mu$m region than
the least luminous quasars.  These differences have the potential for
being diagnostics for the physical parameters (such as orientation and
dust temperature) that go into AGN dust models
\markcite{nie02,dvb05}(e.g., {Nenkova}, {Ivezi{\' c}}, \&  {Elitzur} 2002; {Dullemond} \& {van Bemmel} 2005).  {\em Spitzer}-IRS spectra are needed to
determine if these are truly broad features or if perhaps they are due
to emission lines (e.g., Pa$\alpha$ and Br$\alpha$, which are at
4.05$\mu$m and 1.88$\mu$m, respectively).  In the next section, we
will investigate the differences between these SEDs in terms of the
bolometric corrections derived from them.

\section{Bolometric Luminosities and Accretion Rates}
\label{sec:bol}

The determinations of quasar physical parameters such as bolometric
luminosity, black hole mass, and accretion rate have been
revolutionized by two bodies of work from the past decade.  The first
is the construction of mean quasars SED such as by \markcite{ewm+94}{Elvis} {et~al.} (1994),
which allows one to estimate the bolometric luminosity of a quasar
from a monochromatic luminosity and the assumption of a standard SED.
The other is the reverberation mapping work led by \markcite{wpm99}{Wandel}, {Peterson}, \& {Malkan} (1999) and
\markcite{ksn+00}{Kaspi} {et~al.} (2000) (see also \markcite{kmn+05}{Kaspi} {et~al.} 2005) which allows one to infer a
characteristic radius for the broad emission line region from the
bolometric luminosity of the object and thus to relate the widths of
emission lines to the mass of the central black hole.  With these two
pieces of information, we can derive a third (related) quasar property
--- the Eddington ratio ($L_{\rm Bol}/L_{\rm Edd}$, or equivalently
$L_{\rm bol}/M_{\rm BH}$), which is a measure of the accretion rate.

Our primary purpose for constructing these new SEDs is to investigate
how the bolometric luminosities and accretion rates derived from them
differ from those determined by assuming the mean SED from
\markcite{ewm+94}{Elvis} {et~al.} (1994).  As discussed above, the biases inherent to the sample
of objects used by \markcite{ewm+94}{Elvis} {et~al.} (1994) in addition to the warnings by
\markcite{ewm+94}{Elvis} {et~al.} (1994) of the diversity of individual SEDs, coupled with the
use of their mean SED as a single universal template, is what
motivates this investigation.

\begin{figure}[p]
\epsscale{1.0}
%\plotone{figures/bc.ps}
\plotone{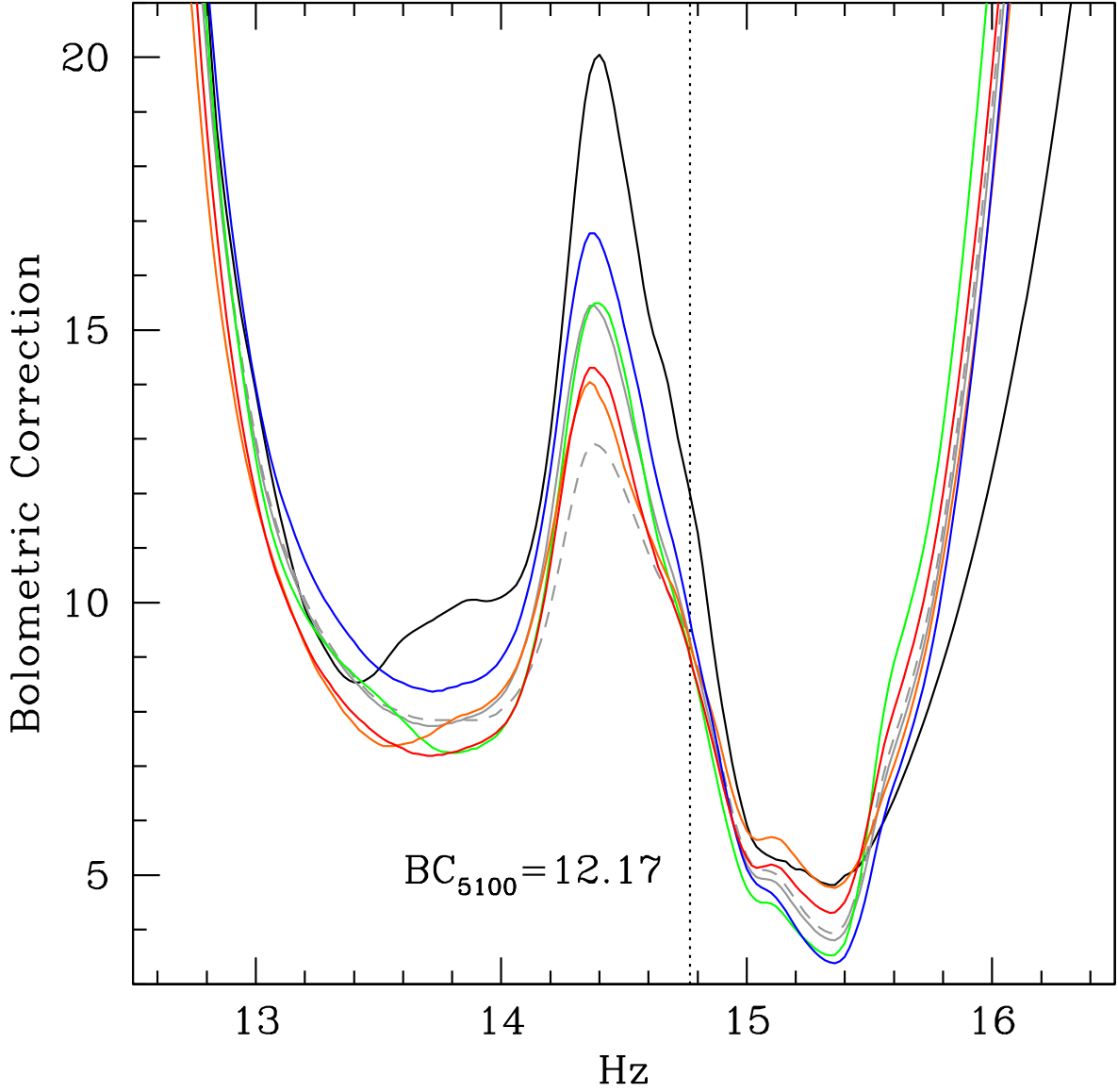}
\caption{Bolometric corrections as a function of frequency for the
SEDs in Fig.~\ref{fig:fig11}, the color coding is the same as for the
previous figure (except that the full SDSS-DR3 mean quasar SED is now
shown in gray instead of cyan).  The dashed gray curve shows the
full-sample mean SED without correction for the host galaxy
contribution.  The bolometric correction from 5100\AA\ for the
\markcite{ewm+94}{Elvis} {et~al.} (1994) radio-quiet SED is 12.17 (dotted line).
\label{fig:fig12}}
\end{figure}

Figure~\ref{fig:fig12} shows the frequency-dependent bolometric
corrections (where the bolometric luminosity is taken to be the
100$\mu$m to 10\,keV integrated luminosity) derived from each of the
mean SEDs shown in Figure~\ref{fig:fig11}.  The bolometric correction
from 5100\AA\ for the \markcite{ewm+94}{Elvis} {et~al.} (1994) radio-quiet SED is
12.17.\footnote{The 5100\AA\ bolometric correction values between 8
and 10 are more commonly used in the literature
\markcite{ksn+00}(e.g., {Kaspi} {et~al.} 2000).}  
%This figure emphasizes what we saw from the
%previous figure, which is that relative to the optical the
%\markcite{ewm+94}{Elvis} {et~al.} (1994) SED over-emphasizes the X-ray and under-emphasizes the
%mid-IR, although the 5100\AA\ bolometric correction remains similar.
The scale in Figure~\ref{fig:fig12} (compared to that in
Fig.~\ref{fig:fig11}) further highlights the differences between the
mean SED shapes.  As our SDSS-selected sample is less biased than that
of \markcite{ewm+94}{Elvis} {et~al.} (1994), it should provide a more robust template for type 1
quasars, and further, the larger sample of objects allows us to
consider the quasar SED not only in the mean, but as a function of
various quasar properties.  We see that in the MIR in particular, it
is important to consider the range of SEDs possible when converting a
monochromatic MIR luminosity to a bolometric luminosity.  Curiously,
in the rest-frame optical, where bolometric corrections are normally
determined (usually at 5100\AA), the differences in the composite SEDs
are quite minimal.  It may be that this lack of a difference results
from this region being a relative minimum in the combination of host
galaxy contamination in the near-IR and dust extinction in the UV.

\begin{figure}[p]
\epsscale{1.0}
%\plotone{figures/bchist.ps}
\plotone{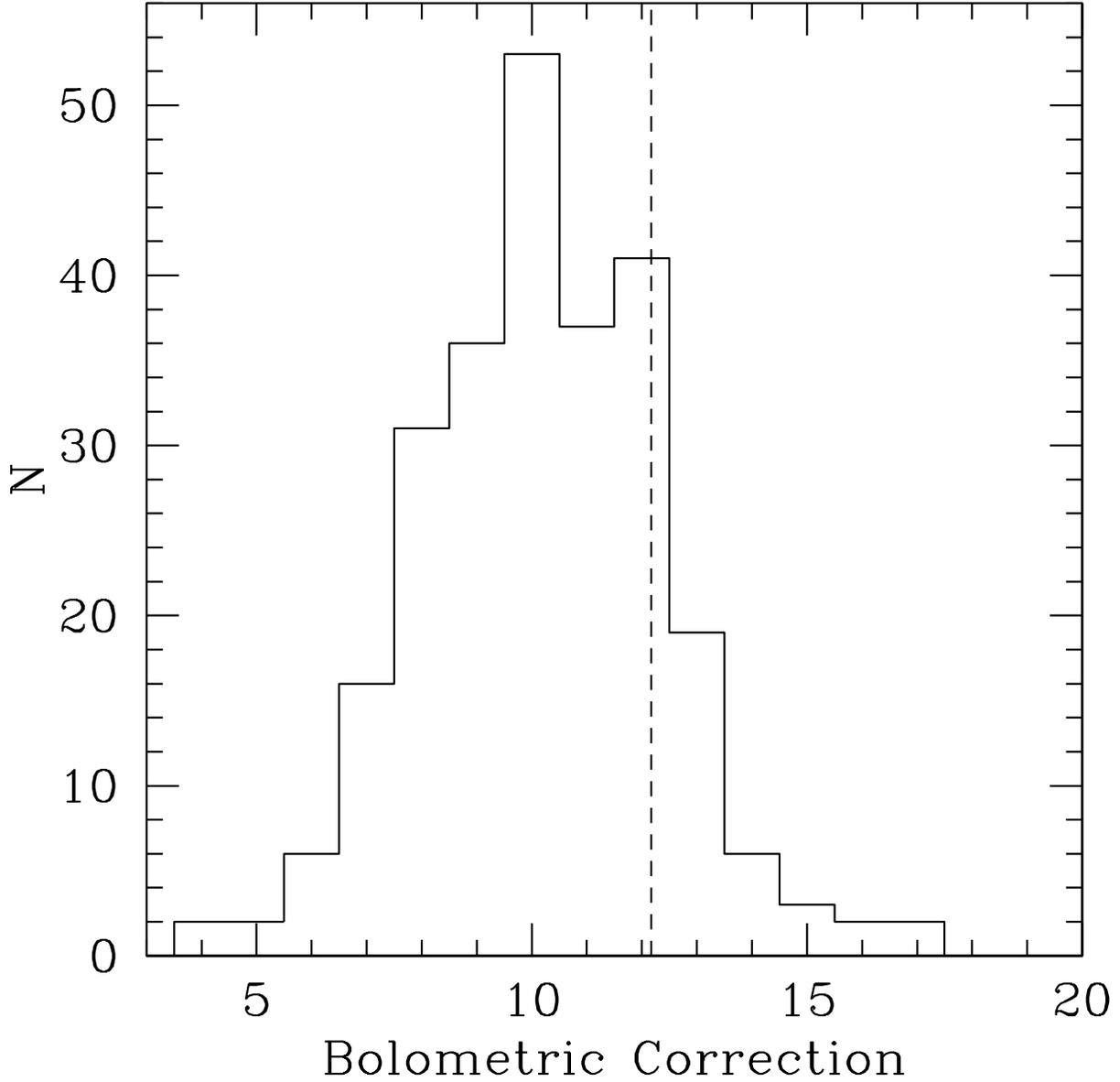}
\caption{Histogram of bolometric corrections (from 5100\AA\ to 100$\mu
m$--10\,keV) for our sample.  According to our data, the bolometric
correction (and thus the accretion rate) is overestimated by 17\% on
the average if the \markcite{ewm+94}{Elvis} {et~al.} (1994) RQ SED is assumed to apply to all
quasars.  Individual objects can have their bolometric corrections
mis-estimated by up to a factor of 2 when assuming a mean SED.
\label{fig:fig13}}
\end{figure}

Looking at the mean SEDs underscores the differences for individual
objects, however.  Figure~\ref{fig:fig13} shows a histogram of the
individual bolometric corrections for each quasar in our sample.
These values (and $L_{\rm bol}$) are also given in
Table~\ref{tab:tab1}.  The mean B-band bolometric correction from
\markcite{ewm+94}{Elvis} {et~al.} (1994) was 11.8 with a range of [5.5,24.7]; our mean and
standard deviation are $10.4\pm2.5$.  In this figure, we see that the
average quasar has its bolometric luminosity overestimated by
$\sim$17\% when using the \markcite{ewm+94}{Elvis} {et~al.} (1994) SED and that objects can have
their bolometric luminosities mis-estimated by up to a factor of 2.
These errors propagate directly to errors in the accretion rate.
Unfortunately, we find no strong trends between the bolometric
correction and color or luminosity.  Thus it is difficult to know when
to apply anything other than the mean bolometric correction.  Clearly,
if we are ever to understand the accretion rate distribution of
quasars, we must attempt to determine bolometric corrections to an
accuracy better than that which is afforded by assuming the mean SED.

A final caveat is that it must be understood that bolometric
corrections and bolometric luminosities determined by summing up all
of the observed flux are really {\em line-of-sight} values that assume
that quasars emit isotropically, when, in fact, we know that this is
not the case.  For example, the same quasar seen both face-on and
edge-on need not yield SEDs that sum to the same bolometric
luminosity; all of the energy must eventually escape, but not
necessarily equally along all lines of sight.  Thus in
Table~\ref{tab:tab1}, in addition to the bolometric luminosity, we
also compute the integrated optical and IR luminosities.  The
integrated IR luminosity may be a more appropriate luminosity measure
if the IR emission is largely isotropic (although even at $30\mu$m, IR
emission is expected to be highly anisotropic; \markcite{vBd03}{van Bemmel} \& {Dullemond} 2003).
Alternatively, one might prefer to use the integrated optical/UV
luminosity since nearly all photons emitted at other wavelengths by
AGNs are reprocessed optical/UV/soft-X-ray seed photons.  In either
case, until we fully understand the SEDs of quasars as a function of
observed quasar properties, and we can associate those properties with
accurate geometrical models, bolometric luminosity and accretion rate
determination will continue to have a significant degree of
uncertainty.

\section{Conclusions}

We have compiled a sample of 259 SDSS type 1 quasars with 4-band {\em
Spitzer}-IRAC detections.  These data are supplemented with
multiwavelength data spanning the radio to X-ray where available.  We
have shown that while MIR-only selection of AGNs is indeed very
efficient, adding optical morphology and color information allows for
even more efficient selection of type 1 quasars.  This multiwavelength
data set was used to construct new mean SEDs of quasars as a function
of color and luminosity, such as are needed to span the diversity of
quasars in terms of computing accurate bolometric luminosities and
accretion rates.  It was shown that computing a bolometric luminosity
by assuming a single mean quasar SED can lead to errors as large as a
factor of two, which translates directly to an error in the presumed
accretion rate.  Finally, judicious use of the knowledge of the mean
quasar SED makes it possible to estimate the fraction of AGNs that are
obscured in the optical by considering the redshift dependence of the
relative flux limits between the optical and MIR.

\acknowledgements

Funding for the SDSS and SDSS-II has been provided by the Alfred
P. Sloan Foundation, the Participating Institutions, the National
Science Foundation, the U.S. Department of Energy, the National
Aeronautics and Space Administration, the Japanese Monbukagakusho, the
Max Planck Society, and the Higher Education Funding Council for
England. The SDSS Web Site is http://www.sdss.org/.  The SDSS is
managed by the Astrophysical Research Consortium for the Participating
Institutions. The Participating Institutions are the American Museum
of Natural History, Astrophysical Institute Potsdam, University of
Basel, Cambridge University, Case Western Reserve University,
University of Chicago, Drexel University, Fermilab, the Institute for
Advanced Study, the Japan Participation Group, Johns Hopkins
University, the Joint Institute for Nuclear Astrophysics, the Kavli
Institute for Particle Astrophysics and Cosmology, the Korean
Scientist Group, the Chinese Academy of Sciences (LAMOST), Los Alamos
National Laboratory, the Max-Planck-Institute for Astronomy (MPA), the
Max-Planck-Institute for Astrophysics (MPIA), New Mexico State
University, Ohio State University, University of Pittsburgh,
University of Portsmouth, Princeton University, the United States
Naval Observatory, and the University of Washington.  This work is
based on observations made with the {\em Spitzer Space Telescope},
operated by the Jet Propulsion Laboratory (JPL), California Institute
of Technology under contract with NASA.  Support was provided by NASA
through JPL.  GTR further acknowledges support from a Gordon and Betty
Moore Fellowship in data intensive sciences.  We thank Eilat Glikman
for making her IR quasar composite spectrum available in advance of
publication.

\appendix
\section*{Appendix}

\begin{figure}[p]
\epsscale{1.0}
%\plotone{figures/sdssspec1.ps}
\plotone{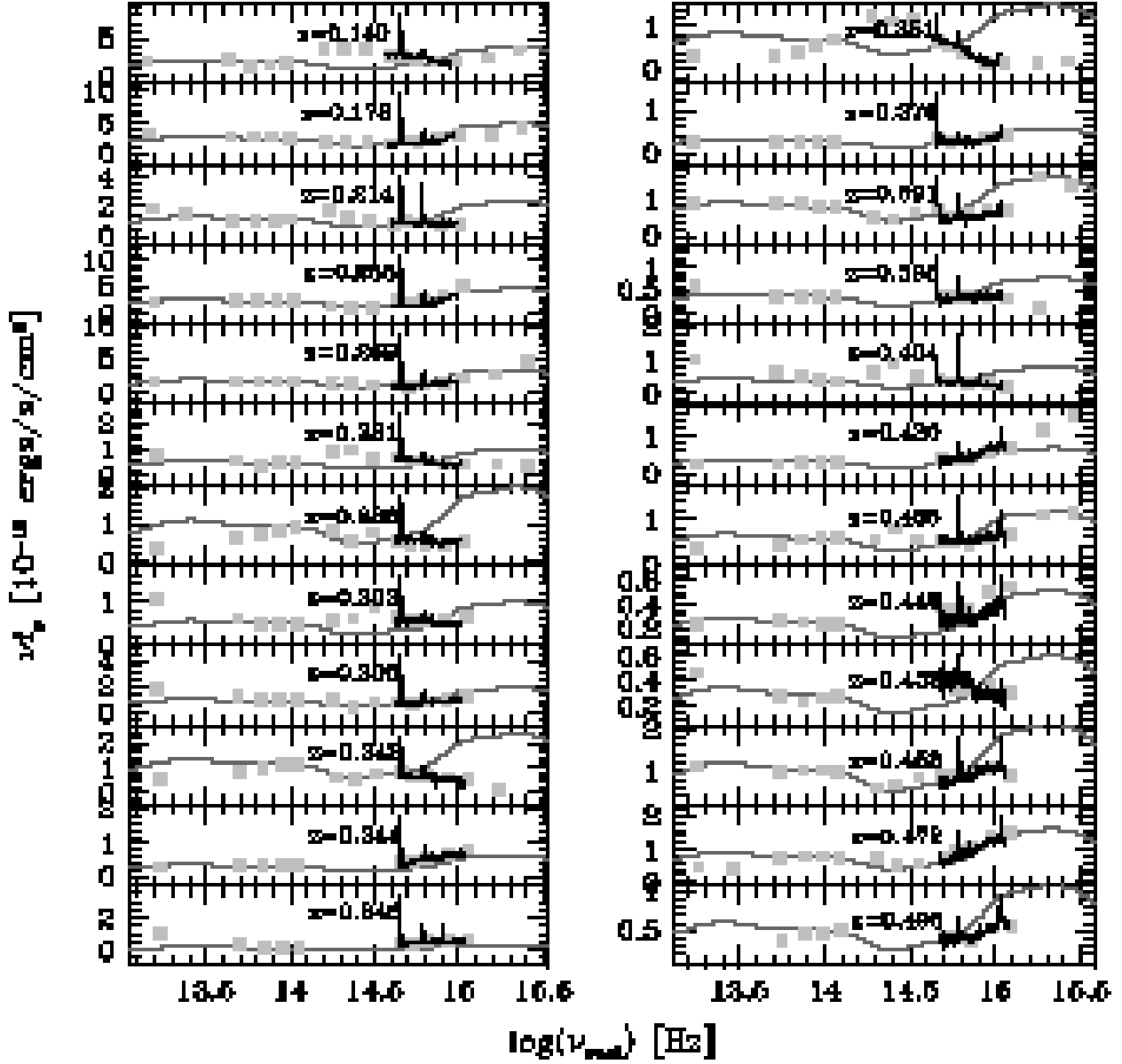}
\caption{Individual mid-IR through UV SEDs for the quasars in our
sample.  The objects are ordered and labeled by redshift to correspond
with the entries in Tables~\ref{tab:tab1} and \ref{tab:tab2}.  The
black curve shows the SDSS spectrum (scaled by the average magnitude
difference between the $g$-, $r$-, and $i$-band point-spread-function
and fiber magnitudes to correct for fiber losses).  The gray curve is
the \markcite{ewm+94}{Elvis} {et~al.} (1994) mean radio-quiet SED (normalized to the 3.6$\mu$m
point).  The lighter gray symbols are the available broad-band
photometric measurements for these objects from {\em Spitzer}-MIPS,
{\em ISO}, {\em Spitzer}-IRAC, 2MASS, SDSS, and {\em GALEX}.  No host
galaxy correction has been applied.
\label{fig:fig14}}
\end{figure}

\begin{figure}[p]
\epsscale{1.0}
%\plotone{figures/sdssspec2.ps}
\plotone{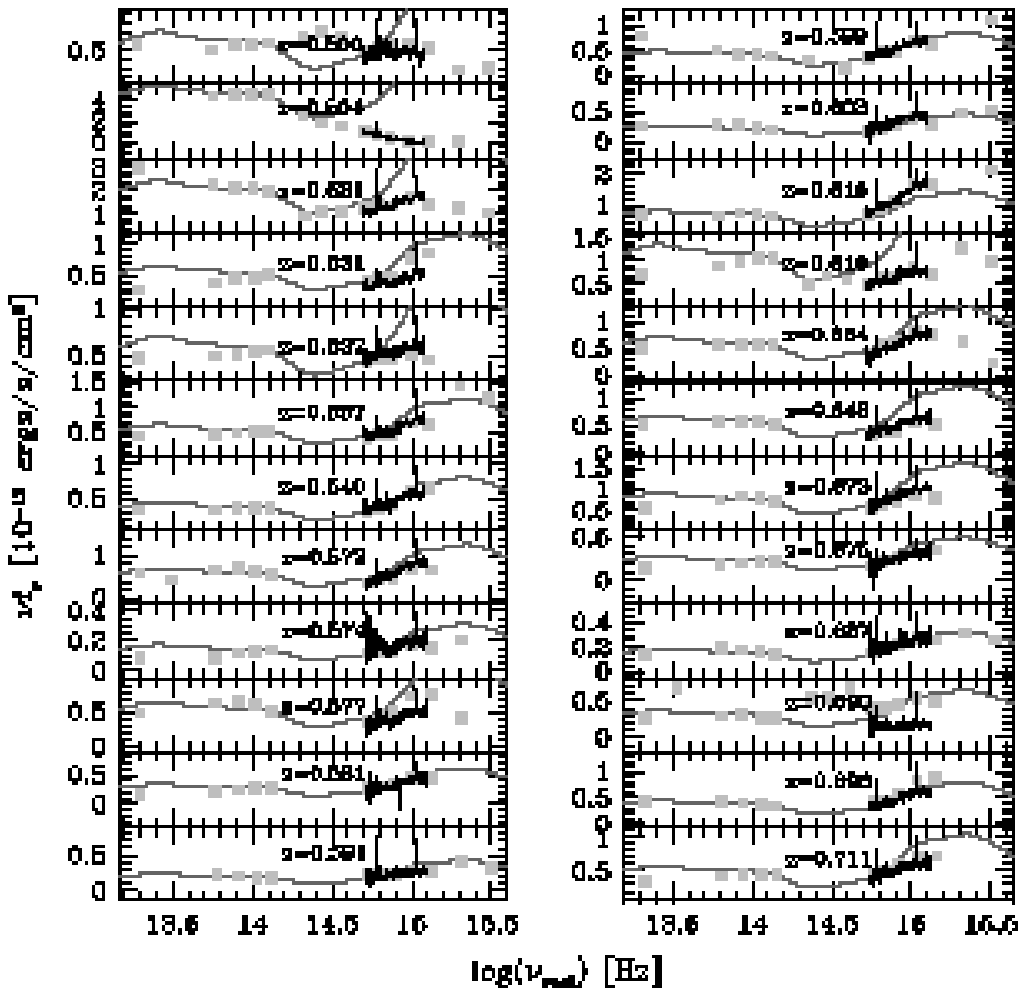}
\caption{Same as for Fig.~\ref{fig:fig14}.
\label{fig:fig15}}
\end{figure}

\begin{figure}[p]
\epsscale{1.0}
%\plotone{figures/sdssspec3.ps}
\plotone{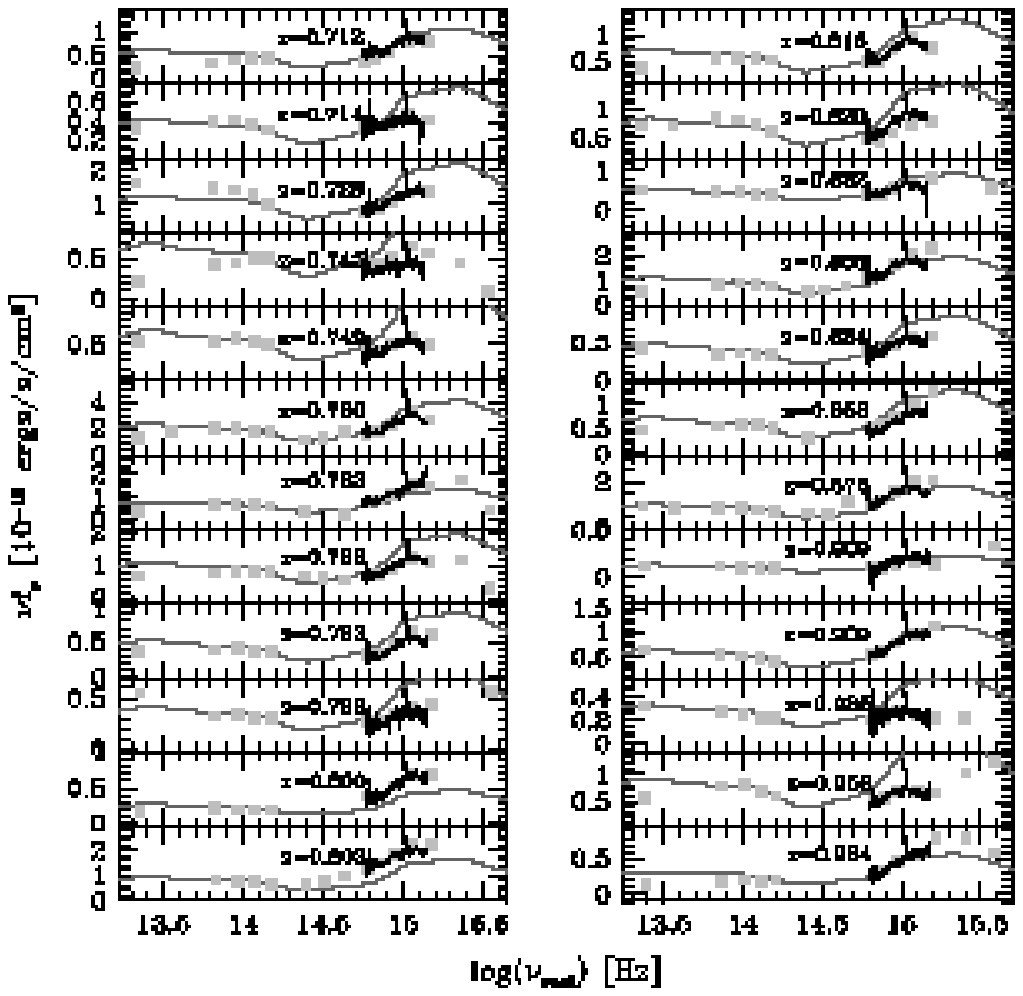}
\caption{Same as for Fig.~\ref{fig:fig14}.
\label{fig:fig16}}
\end{figure}

\begin{figure}[p]
\epsscale{1.0}
%\plotone{figures/sdssspec4.ps}
\plotone{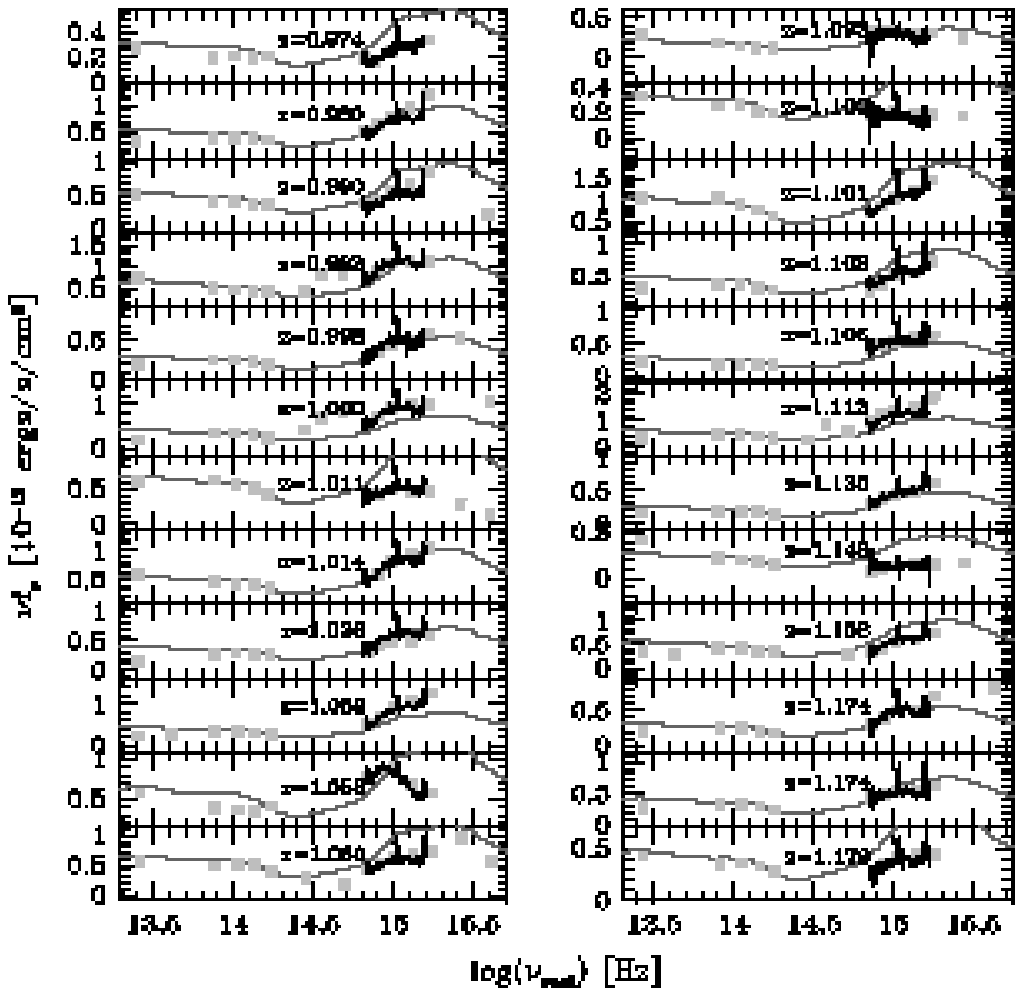}
\caption{Same as for Fig.~\ref{fig:fig14}.
\label{fig:fig17}}
\end{figure}

\begin{figure}[p]
\epsscale{1.0}
%\plotone{figures/sdssspec5.ps}
\plotone{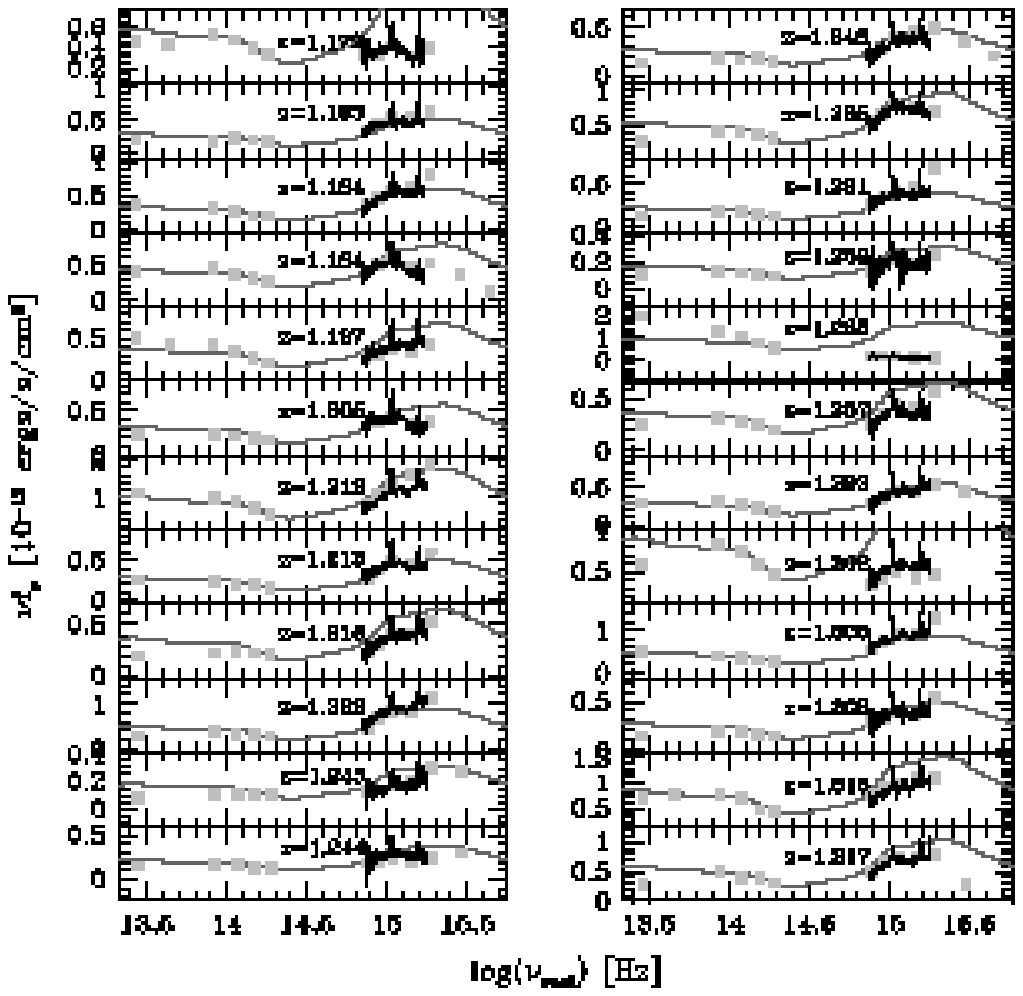}
\caption{Same as for Fig.~\ref{fig:fig14}.
\label{fig:fig18}}
\end{figure}

\begin{figure}[p]
\epsscale{1.0}
%\plotone{figures/sdssspec6.ps}
\plotone{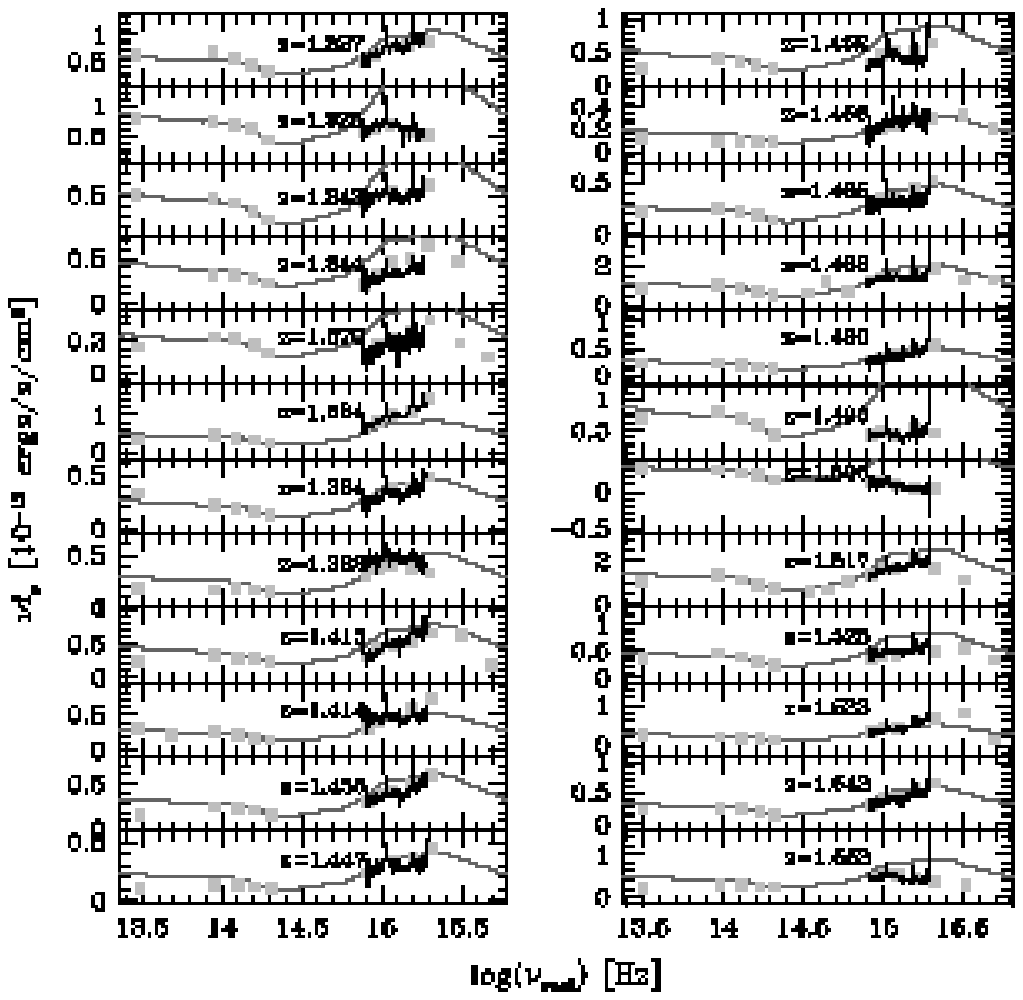}
\caption{Same as for Fig.~\ref{fig:fig14}.
\label{fig:fig19}}
\end{figure}

\begin{figure}[p]
\epsscale{1.0}
%\plotone{figures/sdssspec7.ps}
\plotone{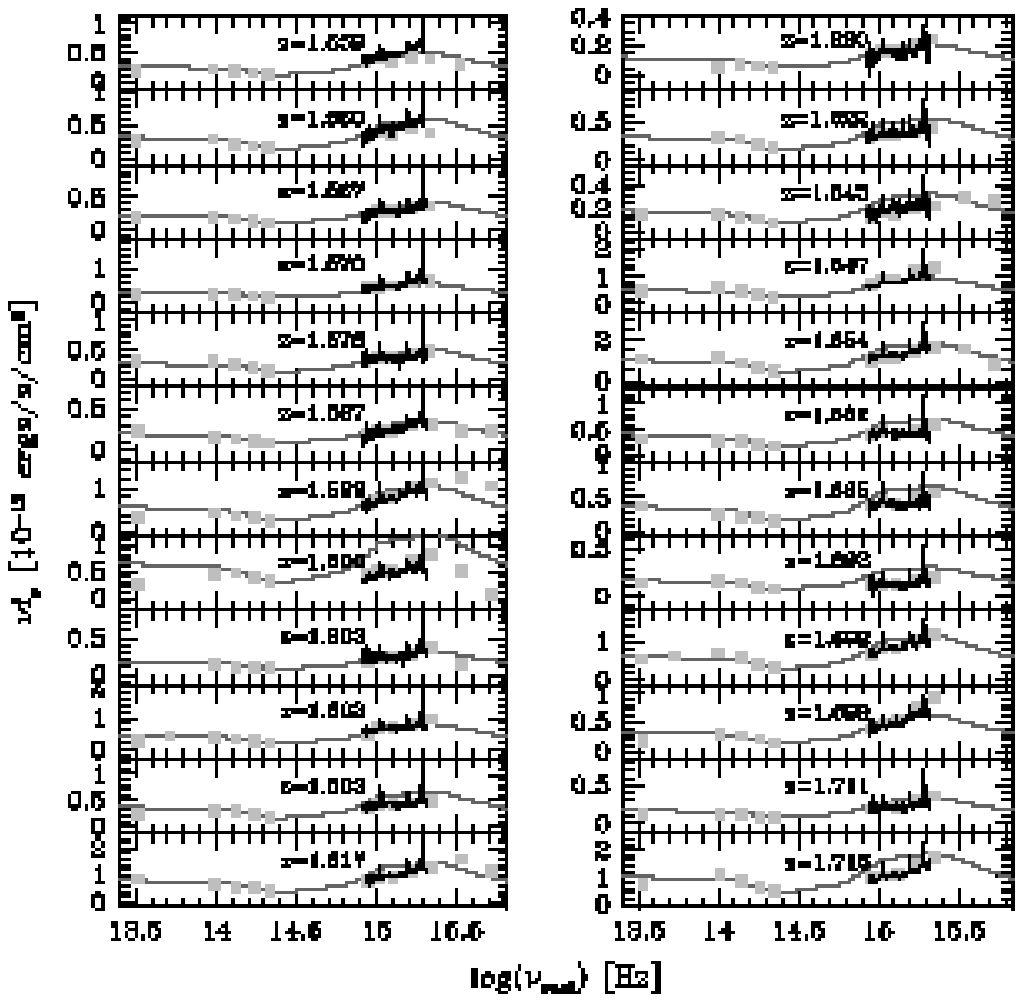}
\caption{Same as for Fig.~\ref{fig:fig14}.
\label{fig:fig20}}
\end{figure}

\begin{figure}[p]
\epsscale{1.0}
%\plotone{figures/sdssspec8.ps}
\plotone{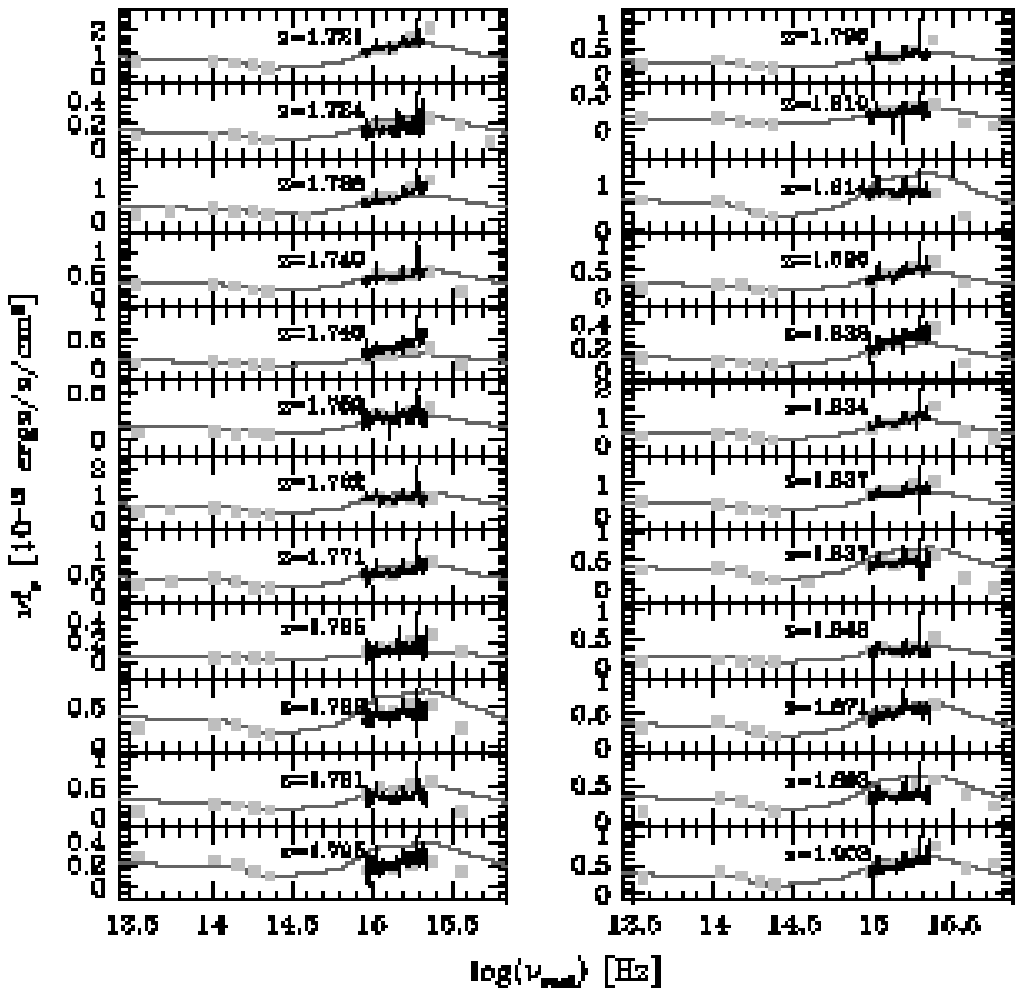}
\caption{Same as for Fig.~\ref{fig:fig14}.
\label{fig:fig21}}
\end{figure}

\begin{figure}[p]
\epsscale{1.0}
%\plotone{figures/sdssspec9.ps}
\plotone{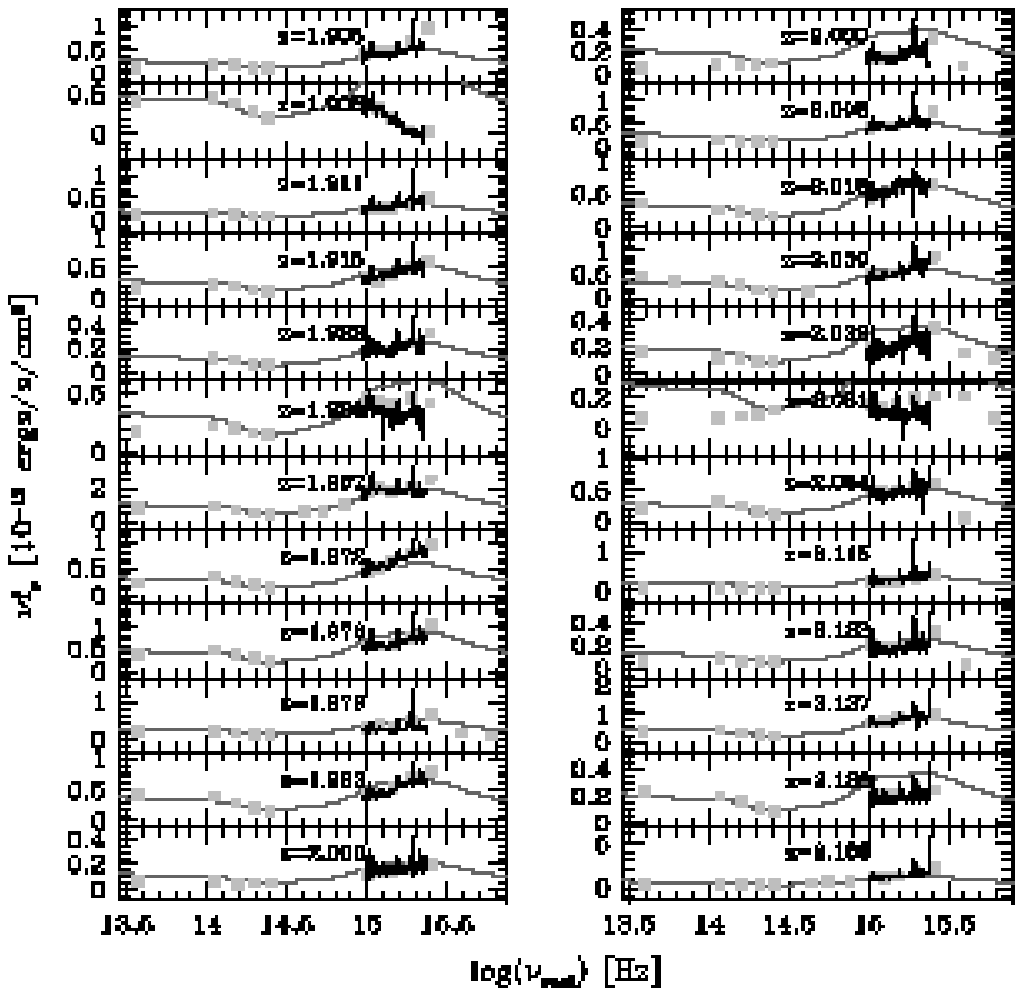}
\caption{Same as for Fig.~\ref{fig:fig14}.
\label{fig:fig22}}
\end{figure}

\begin{figure}[p]
\epsscale{1.0}
%\plotone{figures/sdssspec10.ps}
\plotone{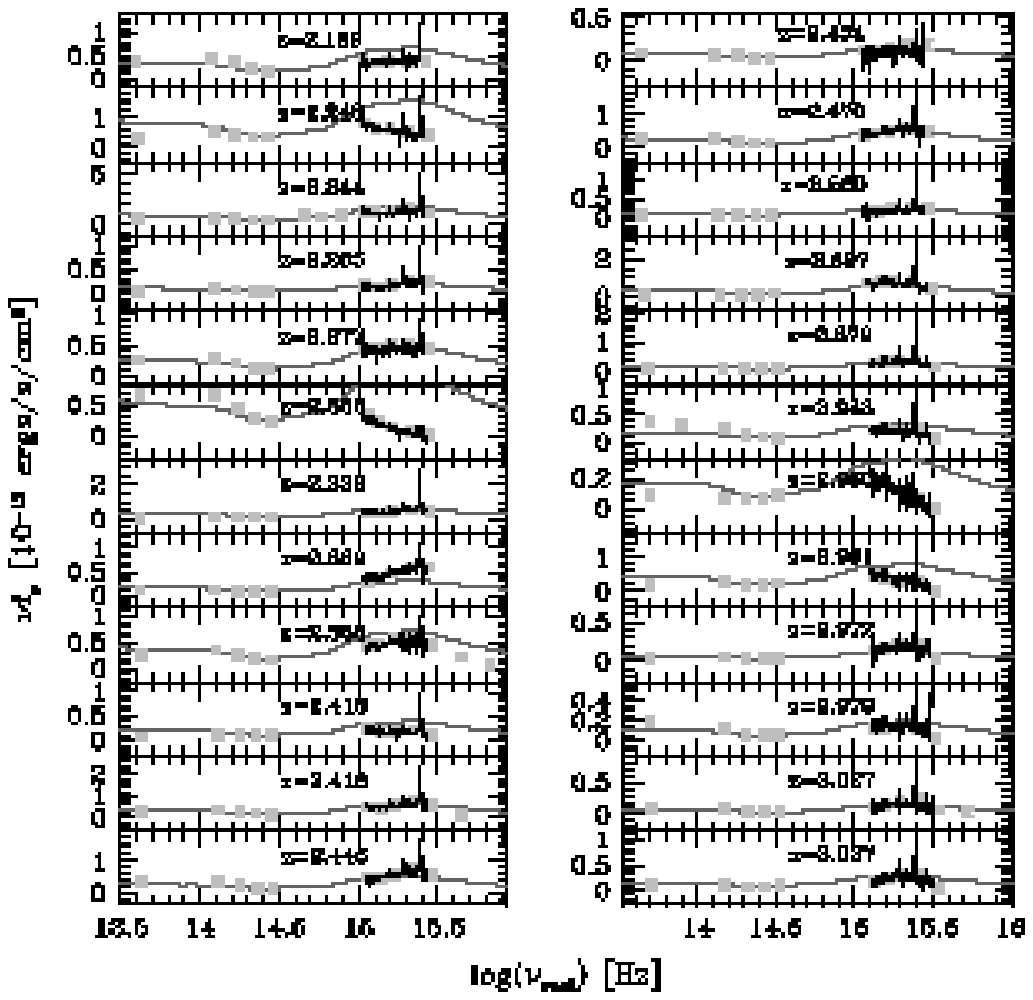}
\caption{Same as for Fig.~\ref{fig:fig14}.
\label{fig:fig23}}
\end{figure}

\begin{figure}[p]
\epsscale{1.0}
%\plotone{figures/sdssspec11.ps}
\plotone{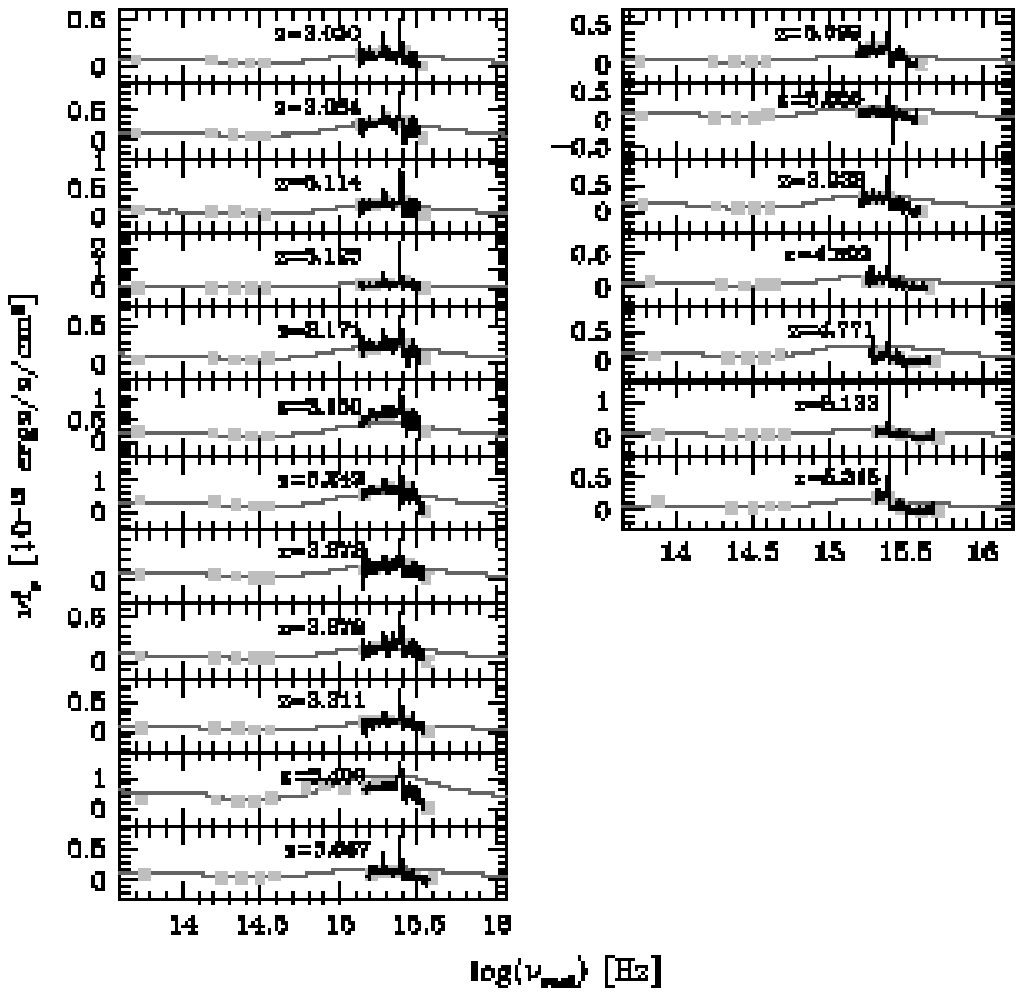}
\caption{Same as for Fig.~\ref{fig:fig14}.
\label{fig:fig24}}
\end{figure}

Figures~\ref{fig:fig14} through \ref{fig:fig24} present the individual
SEDs of each of the 259 quasars in our sample.  The plots are ordered
by redshift to match Tables~\ref{tab:tab1} and \ref{tab:tab2}.  The
individual data points are shown by gray squares.  These are the
observed values; host galaxy contamination has {\em not} been removed.
The SDSS spectra are shown as solid black lines (smoothed by a 15
pixel boxcar).  The SDSS spectra have been scaled by the difference
between the $g$, $r$, and $i$ band PSF and fiber (3\farcs0) magnitudes
to account for light losses due to the finite size of the optical
fibers.  The \markcite{ewm+94}{Elvis} {et~al.} (1994) radio-quiet mean SED is shown by the gray
curves and is normalized to the 3.6$\mu$m data point of each quasar.

\clearpage

\nocite{Surace05}
\nocite{Fadda06}

%% \bibliography

%\bibliographystyle{apj}

\clearpage

\clearpage

\clearpage

\clearpage

\begin{deluxetable}{lllllrllllllll}
\rotate
\tabletypesize{\tiny}
\tablewidth{0pt}
\tablecaption{SDSS-{\em Spitzer} Quasar Photometry I\label{tab:tab1}}
\tablehead{
\colhead{Name} &
\colhead{$z_{em}$} &
\colhead{$L_{\rm bol}$\tablenotemark{a}} &
\colhead{$L_{\rm opt}$\tablenotemark{b}} &
\colhead{$L_{\rm ir}$\tablenotemark{c}} &
\colhead{BC\tablenotemark{a}} &
\colhead{X-ray} &
\colhead{$f$} &
\colhead{$n$} &
\colhead{$u$} &
\colhead{$g$} &
\colhead{$r$} &
\colhead{$i$} &
\colhead{$z$} \\
\colhead{(SDSS J)} &
\colhead{} &
\colhead{lg(ergs/s)} &
\colhead{lg(ergs/s)} &
\colhead{lg(ergs/s)} &
\colhead{} &
\colhead{lg(c/s)} &
\colhead{(AB mag)} &
\colhead{(AB mag)} &
\colhead{(AB mag)} &
\colhead{(AB mag)} &
\colhead{(AB mag)} &
\colhead{(AB mag)} &
\colhead{(AB mag)}
}
\startdata
\input{tab1}
\enddata
\tablenotetext{a}{Bolometric (100$\mu$m to 10\,keV) luminosity and bolometric correction (from 5100\AA).}
\tablenotetext{b}{100$\mu$m to 1$\mu$m integrated luminosity.}
\tablenotetext{c}{1$\mu$m to 0.1$\mu$m integrated luminosity.}
\end{deluxetable}

\begin{deluxetable}{llllcccccccrr}
\rotate
\tabletypesize{\tiny}
\tablewidth{0pt}
\tablecaption{SDSS-{\em Spitzer} Quasar Photometry II\label{tab:tab2}}
\tablehead{
\colhead{Name} &
\colhead{$J$} &
\colhead{$H$} &
\colhead{$K$} &
\colhead{$S_{3.6}$} &
\colhead{$S_{4.5}$} &
\colhead{$S_{5.8}$} &
\colhead{$S_{8.0}$} &
\colhead{$S_{15}$} &
\colhead{$S_{24}$} &
\colhead{$S_{70}$} &
\colhead{radio} &
\colhead{$L_{\rm rad}$} \\
\colhead{(SDSS J)} &
\colhead{(Vega)} &
\colhead{(Vega)} &
\colhead{(Vega)} &
\colhead{($\mu$Jy)} &
\colhead{($\mu$Jy)} &
\colhead{($\mu$Jy)} &
\colhead{($\mu$Jy)} &
\colhead{(mJy)} &
\colhead{(mJy)} &
\colhead{(mJy)} &
\colhead{(mJy)} &
\colhead{lg(ergs/s/Hz)}
}
\startdata
\input{tab2}
\enddata
\end{deluxetable}

\begin{deluxetable}{rrrrrrrr}
\tabletypesize{\tiny}
\tablewidth{0pt}
\tablecaption{Mean Quasar SEDs\label{tab:tab3}}
\tablehead{
\colhead{Log Hz} &
\colhead{All} &
\colhead{Blue} &
\colhead{Red} &
\colhead{OptLum} &
\colhead{OptDim} &
\colhead{IRLum} &
\colhead{IRDim}
}
\startdata
\input{tab3}
\enddata
\tablecomments{All of the SEDs are taken to have $\alpha_x=-1.0$ beyond $10^{17}$\,Hz and thus have constant luminosity beyond this value.  Units are $\log({\rm ergs\; s^{-1}})$.}
\end{deluxetable}

\end{document}

%% file: tab1.tex
105705.39+580437.4 & 0.140 & 45.06 & 44.49 & 44.78 & 10.60 & -0.708 & 18.31$\pm$0.08 & 18.15$\pm$0.04 & 17.92$\pm$0.03 & 17.61$\pm$0.05 & 17.25$\pm$0.02 & 16.83$\pm$0.02 & 16.56$\pm$0.04 \\
171902.28+593715.9 & 0.178 & 45.21 & 44.74 & 44.93 & 9.41 & -1.221 & 18.10$\pm$0.01 & 17.99$\pm$0.01 & 17.49$\pm$0.02 & 17.50$\pm$0.02 & 17.36$\pm$0.02 & 17.06$\pm$0.02 & 17.20$\pm$0.02 \\
160655.34+534016.8 & 0.214 & 45.13 & 44.45 & 44.91 & 11.87 & $\ldots$ & $\ldots$ & $\ldots$ & 18.85$\pm$0.03 & 18.71$\pm$0.02 & 18.22$\pm$0.02 & 17.86$\pm$0.02 & 17.91$\pm$0.03 \\
163111.28+404805.2 & 0.258 & 45.68 & 45.27 & 45.19 & 9.84 & -0.551 & $\ldots$ & $\ldots$ & 16.98$\pm$0.01 & 17.05$\pm$0.02 & 17.08$\pm$0.01 & 17.10$\pm$0.01 & 16.86$\pm$0.01 \\
171207.44+584754.4 & 0.269 & 45.49 & 45.06 & 45.12 & 12.29 & -1.235 & 17.97$\pm$0.01 & 18.08$\pm$0.01 & 17.83$\pm$0.02 & 17.93$\pm$0.02 & 17.88$\pm$0.02 & 17.94$\pm$0.02 & 17.51$\pm$0.02 \\
171033.21+584456.8 & 0.281 & 45.15 & 44.46 & 44.95 & 10.44 & $\ldots$ & 20.59$\pm$0.04 & 20.06$\pm$0.02 & 19.58$\pm$0.03 & 19.25$\pm$0.03 & 18.70$\pm$0.02 & 18.52$\pm$0.02 & 18.10$\pm$0.03 \\
105644.52+572233.4 & 0.286 & 45.08 & 44.53 & 44.78 & 10.12 & $\ldots$ & $\ldots$ & $\ldots$ & 19.36$\pm$0.03 & 19.26$\pm$0.02 & 18.88$\pm$0.02 & 18.69$\pm$0.02 & 18.32$\pm$0.02 \\
104739.49+563507.2 & 0.303 & 45.19 & 44.66 & 44.88 & 9.82 & $\ldots$ & $\ldots$ & $\ldots$ & 19.16$\pm$0.04 & 19.02$\pm$0.04 & 18.72$\pm$0.04 & 18.57$\pm$0.03 & 18.20$\pm$0.03 \\

%% file: tab2.tex
105705.39+580437.4 & 14.99$\pm$0.08 & 14.21$\pm$0.09 & 13.48$\pm$0.07 & 2351.5$\pm$5.6 & 2366.9$\pm$7.3 & 2838.0$\pm$15.1 & 6273.4$\pm$16.1 & $\ldots$ & 16.61$\pm$0.02 & 98.0$\pm$0.6 & $\ldots$ & $<$29.69\\
171902.28+593715.9 & 15.89$\pm$0.09 & 15.02$\pm$0.09 & 14.15$\pm$0.06 & 2925.1$\pm$293.1 & 4095.1$\pm$409.8 & 5365.6$\pm$541.1 & 7193.8$\pm$720.4 & $\ldots$ & 26.91$\pm$0.04 & 22.9$\pm$4.0 & 0.23 & 29.28\\
160655.34+534016.8 & 16.37$\pm$0.10 & 15.33$\pm$0.11 & 14.32$\pm$0.07 & 1396.7$\pm$4.6 & 1657.2$\pm$5.7 & 2047.6$\pm$14.1 & 2973.1$\pm$10.9 & 7.72 & 14.80$\pm$0.02 & 37.7$\pm$1.6 & $\ldots$ & $<$30.09\\
163111.28+404805.2 & 16.24$\pm$0.10 & 15.45$\pm$0.12 & 14.48$\pm$0.09 & 2729.7$\pm$5.0 & 3632.2$\pm$6.6 & 4686.4$\pm$15.1 & 6218.9$\pm$11.9 & $\ldots$ & 16.90$\pm$0.03 & $\ldots$ & $\ldots$ & $<$30.26\\
171207.44+584754.4 & 16.26$\pm$0.10 & 15.36$\pm$0.09 & 14.61$\pm$0.10 & 2024.6$\pm$203.1 & 2411.9$\pm$242.2 & 3162.9$\pm$321.5 & 4353.7$\pm$437.9 & $\ldots$ & 13.34$\pm$0.07 & $\ldots$ & 0.14 & 29.45\\
171033.21+584456.8 & 16.90$\pm$0.18 & 15.68$\pm$0.11 & 14.96$\pm$0.10 & 589.2$\pm$59.9 & 708.4$\pm$71.8 & 709.8$\pm$78.1 & 1571.6$\pm$159.5 & $\ldots$ & 6.06$\pm$0.07 & 44.0$\pm$8.0 & $\ldots$ & $<$30.34\\
105644.52+572233.4 & 16.71$\pm$0.10 & 16.33$\pm$0.27 & 15.10$\pm$0.12 & 1161.4$\pm$4.4 & 1280.1$\pm$5.2 & 1417.1$\pm$13.2 & 1742.5$\pm$9.7 & $\ldots$ & 3.18$\pm$0.02 & $\ldots$ & $\ldots$ & $<$30.36\\
104739.49+563507.2 & 16.63$\pm$0.16 & 16.24$\pm$0.24 & 15.51$\pm$0.18 & 572.2$\pm$2.6 & 671.2$\pm$2.6 & 886.3$\pm$10.7 & 1566.3$\pm$6.5 & $\ldots$ & 8.61$\pm$0.02 & $\ldots$ & $\ldots$ & $<$30.41\\

%% file: tab3.tex
12.50 & 44.43 & 44.39 & 44.47 & 44.76 & 44.12 & 44.82 & 44.10\\
12.52 & 44.49 & 44.44 & 44.53 & 44.81 & 44.18 & 44.87 & 44.16\\
12.54 & 44.54 & 44.50 & 44.58 & 44.87 & 44.23 & 44.93 & 44.22\\
12.56 & 44.59 & 44.55 & 44.64 & 44.93 & 44.28 & 44.98 & 44.27\\
12.58 & 44.65 & 44.60 & 44.69 & 44.98 & 44.33 & 45.04 & 44.32\\
12.60 & 44.70 & 44.65 & 44.74 & 45.03 & 44.38 & 45.09 & 44.36\\
12.62 & 44.74 & 44.70 & 44.78 & 45.09 & 44.42 & 45.15 & 44.40\\
12.64 & 44.79 & 44.74 & 44.83 & 45.14 & 44.45 & 45.20 & 44.44\\